\documentclass[preprint]{aastex}

\def\lsim{\mathrel{\rlap{\lower4pt\hbox{\hskip1pt$\sim$}}
    \raise1pt\hbox{$<$}}}         
\def\gsim{\mathrel{\rlap{\lower4pt\hbox{\hskip1pt$\sim$}}
    \raise1pt\hbox{$>$}}}         

\usepackage{natbib}
\usepackage{graphicx}
\pdfoutput=1

\begin{document}


\title{CN-Cycle Solar Neutrinos and Sun's Primordial Core Metalicity}


\author{W. C. Haxton}
\affil{Institute  for   Nuclear  Theory  and  Dept.   Physics,  University  of
  Washington, Seattle, WA 98195, USA}
\email{haxton@u.washington.edu}
\and
\author{A. M. Serenelli}
\affil{Institute for Advanced Study, Princeton, NJ 08540, USA}
\email{aldos@ias.edu}



\begin{abstract}
We argue that it may be possible to exploit neutrinos from the CN cycle and pp
chain  to determine the  primordial solar  core abundances  of C  and N  at an
interesting level of  precision.  Such a measurement would  allow a comparison
of  the  Sun's  deep interior  composition  with  it  surface, testing  a  key
assumption of the standard solar  model (SSM), a homogeneous zero-age Sun.  It
would   also  provide   a   cross-check  on   recent  photospheric   abundance
determinations that have altered the  once excellent agreement between the SSM
and  helioseismology.   As  further   motivation,  we  discuss  a  speculative
possibility   in  which   photospheric  abundance/helioseismology   puzzle  is
connected  with  the   solar-system  metal  differentiation  that  accompanied
formation of the gaseous giant planets.

The  theoretical  relationship between  core  C and  N  and  the $^{13}$N  and
$^{15}$O solar neutrino fluxes can be  made more precise (and more general) by
making use of the Super-Kamiokande and SNO $^8$B neutrino capture rates, which
calibrate  the  temperature  of  the  solar  core.  The  primordial  C  and  N
abundances can then be obtained from  these neutrino fluxes and from a product
of nuclear  rates, with little  residual solar model dependence.   We describe
some of the  recent experimental advances that could  allow this comparison to
be  made  (theoretically)  at  the  $\sim$  9\%  level,  and  note  that  this
uncertainty may  be reduced further  due to ongoing  work on the  S-factor for
$^{14}$N(p,$\gamma$).  The  envisioned measurement might be  possible in deep,
large-volume detectors using organic scintillator, e.g., Borexino or SNO+.
\end{abstract}



\section{Introduction}

Over  three  decades  one of  the  most  intriguing  problems in  physics  and
astrophysics has  been that  of the missing  solar neutrinos,  the discrepancy
between        Ray        Davis's        chlorine-detector        measurements
(\citealt{davis,davis_bahc}) and the 
predictions  of  the standard  solar  model  (SSM)  developed by  Bahcall  and
collaborators \citep{bpb00,bp04,bsb_mc}, by Turck-Chieze and collaborators
\citep{brun1,brun2}, and  others.  Part  of the problem's  fascination has
been the tension between stellar theory and particle physics: arguments for new
neutrino physics required one to believe that {\it ab initio} models correctly
predicted the solar core temperature to an accuracy of about 1\%.

Gradually  the   combination  of  quantitative  tests  of   the  solar  model,
particularly determinations  of the interior sound  speed via helioseismology,
and new solar neutrino experiments, Kamioka \citep{kamioka}, SAGE \citep{sage} 
and GALLEX/GNO \citep{gallex}, and Super-Kamiokande \citep{sk1,sk2}, made the 
arguments for  new physics compelling.   The pattern of solar  neutrino fluxes
proved difficult to attribute to any plausible variation in the SSM.  With the
direct detection  of both  electron- and heavy-flavor  solar neutrinos  in the
Sudbury  Neutrino  Observatory,  solar   neutrino  experiments  had  not  only
demonstrated that electron neutrinos oscillate  on their way to the earth, but
also     determined    the     parameters    governing     that    oscillation
\citep{sno1,sno2,sno3}.  
Progress has continued with  the KamLAND \citep{kamland} reactor experiment and
the  Borexino collaboration's  efforts \citep{borexino}  to  measure low-energy
solar neutrinos in real time.   Borexino's recent results for the $^7$Be solar
neutrino flux are consistent with Large Mixing Angle (LMA) solution.

Because the incorporation of neutrino  mass and mixing into the standard model
requires  new physics,  the  field's  attention has  naturally  turned to  the
unresolved particle physics questions, such as the mass hierarchy, mass scale,
third mixing  angle, and CP-violating phases. \citet{APS}  summarizes the open
questions and the envisioned future experimental program.  

Here we return  to one of the initial motivations  for solar neutrino physics,
using the neutrino flux  as a probe of the SSM.  We  will argue that important
tests  of the Sun  and its  initial conditions  can be  made by  measuring the
CN-cycle neutrinos.  Such a measurement would test a key assumption of the SSM
-- that  convective mixing  during the  early pre-main-sequence  Hayashi phase
produced a homogeneous Sun, and  that subsequent evolution has not appreciably
altered the  distribution of metals -- an  assumption that may now  be in some
degree of  conflict with  helioseismology.  This assumption  is the  basis for
taking the SSM's  primordial core metal abundances from  today's surface metal
abundances.   We  argue   that  a  series  of  recent   advances  --  SNO  and
Super-Kamiokande  measurements  of  the  $^8$B neutrinos,  new  cross  section
measurements for $^{14}$N(p,$\gamma$) and  for certain pp-chain reactions, and
detector developments such as Borexino  and SNO+, might allow one to determine
the primordial  abundances of  C and  N, with little  dependence on  the solar
model: to  good accuracy,  these abundances can  be derived  from experimental
quantities, namely the CN-cycle neutrino  fluxes, the $^8$B neutrino flux, and
nuclear cross sections and oscillation parameters measured in the laboratory.

Such a  check of the SSM  is of added  interest because recent 3D  modeling of
photospheric absorption  lines has  led to a  downward revision in  the metal
content  of  the solar  convective  zone  \citep{ags05}.  This  significantly
alters the 
once  spectacular  agreement  between  the  SSM  and  helioseismology  in  the
temperature region below the solar convective zone, $\sim$ 2-5 $\times$ 10$^6$
K \citep{bbps05,bsb05,antia,montalban}. In this 
region C, N, O, Ne, and Ar are partially 
ionized  and  particularly  O and  Ne  have  a  significant influence  on  the
radiative opacity. 

A  quantitative comparison between  the Sun's  surface and  core metalicities
could prove  useful in understanding  the chemical evolution of  other gaseous
bodies in  our solar system,  whose interiors are  not as easily  probed.  The
Galileo and Cassini  missions found significant metal enrichments  in the H/He
atmospheres of Jupiter and Saturn, e.g.,  abundances of C and N of $\sim$ four
times solar for Jupiter and $\sim$ 4-8 for Saturn \citep{guillot}.  Planetary 
models that  take account of these  data indicate that the  gaseous giants are
very significant  solar-system metal reservoirs.   We discuss, because  of the
size of these reservoirs and the  time they were created, the possibility that
they  might have  some  connection  with the  current  conflict between  solar
interior   (helioseismology)  and   surface  (photospheric   absorption  line)
abundance determinations.

Finally, just  as the solar  neutrino program to  date has provided  our first
quantitative  test of  the nuclear  astrophysics governing  proton  burning in
low-mass main-sequence stars, a solar  CN-cycle neutrino program would give us
our  first   experimental  constraints  on   the  process  by   which  massive
main-sequence stars burn hydrogen.  The CN  cycle is thought to have driven an
early convective stage  in our Sun, and is also important  to the evolution of
the first generation of massive metal-poor stars, where it turns on only after
carbon has been synthesized by the triple alpha process.

\section{The CNO Bi-Cycle and its Neutrinos}

\begin{figure*}
\includegraphics[width=12cm]{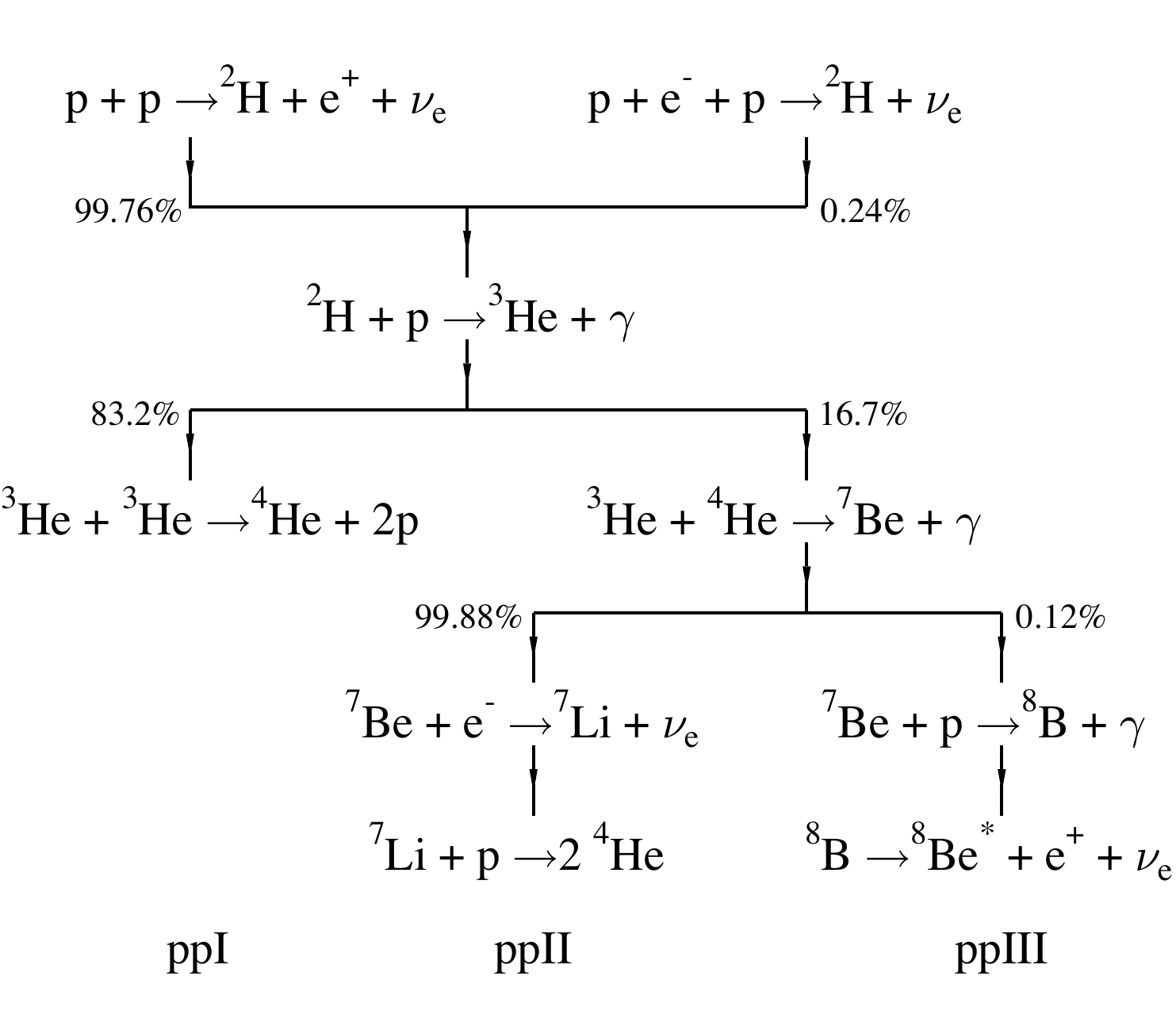}
\caption{ The pp-chain for hydrogen burning.  The relative termination rates
of competing reactions correspond to the BPS08(AGS) SSM. \label{fig:one}} 
\end{figure*}

The need for two mechanisms to  burn hydrogen was recognized in the pioneering
work  of  Bethe  and  collaborators.   The pp-chain,  which  dominates  energy
production  in  our  Sun  and  other  low-mass  main-sequence  stars,  can  be
considered a primary process in  which the chain's ``catalysts'' -- deuterium,
$^3$He, and  $^7$Be/$^7$Li, the  elements participating in  intermediate steps
shown  in  Fig.~\ref{fig:one} --  are  synthesized  as  the chain  burns  to
equilibrium.

\begin{figure*}
\includegraphics[width=12cm]{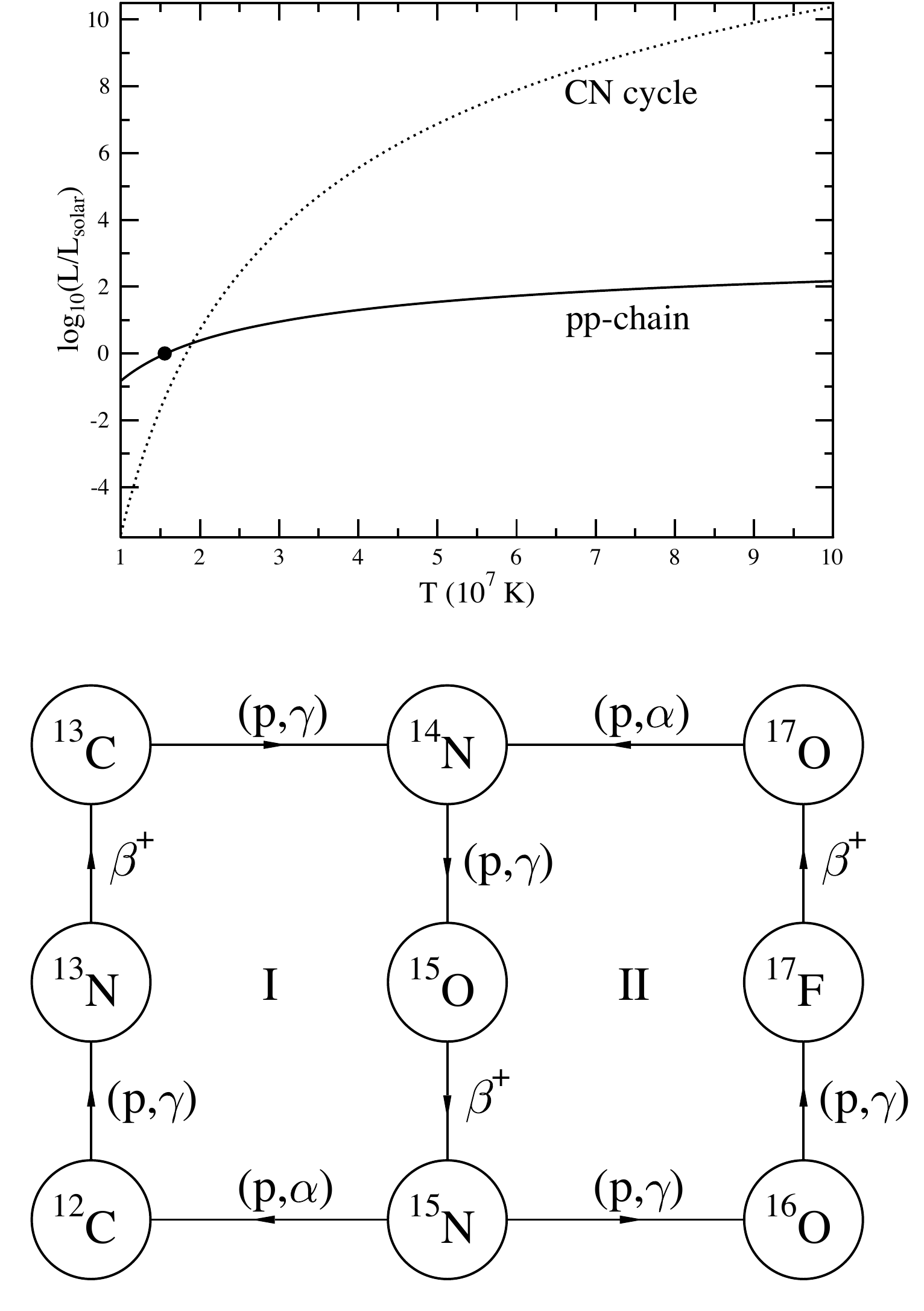}
\caption{The lower panel shows the CNO bi-cycle for hydrogen burning.
The  upper panel  compares  the energy  produced  in the  CN  cycle with  that
produced 
in the pp-chain, as a function of temperature T$_7$, measured
in units of 10$^7$ K.  The results are normalized to the 
pp-chain energy production in the  Sun's central core and to solar metalicity,
and 
assume  the burning  is  in  equilibrium.  The  sharp  CN-cycle dependence  on
temperature 
is apparent.  If approximated as a  power law T$^x$, $x$ ranges between $\sim$
19 
and  $\sim$  22   over  the  range  of  temperatures   typical  of  the  Sun's
hydrogen-burning core. 
The dot marks the point corresponding to the Sun's center,
T$_7 = 1.57$. \label{fig:two}}
\end{figure*}

But the  sharper $T$-dependence of the  CNO cycle is necessary  to account for
the structure  of massive main-sequence  stars.  Unlike the pp-chain,  the CNO
bi-cycle  (Fig.~\ref{fig:two})  is a  secondary process:  the catalysts  for H
burning  are the  pre-existing metals.   Thus the  CNO contribution  to energy
generation is  directly proportional to  the stellar-core number  abundance of
the primordial metals.   The CN-cycle, denoted by I  in Fig.~\ref{fig:two}, is
an important SSM  neutrino source.  The cycle conserves  the number abundance,
but alters the distribution of metals as it burns into equilibrium, eventually
achieving equilibrium abundances proportional to the inverse of the respective
rates.

The reactions controlling early conversion of metals in the solar core and the
approach  to equilibrium  are  $^{12}$C(p,$\gamma$) and  $^{14}$N(p,$\gamma$):
these are  the next-to-slowest and  slowest rates in the  lower-temperature CN
cycle, respectively.  The  central temperature of the solar  core at the onset
of nuclear  burning, $T_7  \sim 1.34$, corresponds  to a $^{12}$C  lifetime of
about  2  $\cdot$ 10$^7$  y.   Thus  the  initial out-of-equilibrium  CN-cycle
conversion of $^{12}$C  to $^{14}$N in the central region of  the early Sun is
complete and  rapid.  The associated energy  release is thought  to render the
central portion of the solar core  convectively unstable for a period of about
$10^8$ y.   That is, the steep temperature  dependence of $^{12}$C(p,$\gamma$)
produces composition, opacity, and  thus thermal gradients sufficient to drive
convection.  The temperature  at which the $^{12}$C lifetime  is comparable to
the Sun's 4.57 b.y.  lifetime is $T_7 \sim 1.0$.  In the SSM this includes
essentially the entire  energy-producing core, $ R \lsim  0.18 R_\odot$ and $M
\lsim 0.29 M_\odot$, so that nearly  all of the core's primordial $^{12}$C has
been converted  to $^{14}$N.  This  change in the chemical  composition alters
the opacity  and, at the 3\%  level, the heavy-element mass  fraction $Z$, SSM
effects first explored by \citet{bu88}. 

The  $^{14}$N(p,$\gamma$)  reaction determines  whether  equilibrium has  been
achieved.  The $^{14}$N  lifetime is shorter than the age of  the Sun for $T_7
\gsim 1.33$.  Therefore equilibrium for the CN cycle has been reached only for
$R \lsim  0.1 R_\odot$, corresponding to the  central 7\% of the  Sun by mass.
Consequently, over a significant portion  of the outer core, $^{12}$C has been
converted   to  $^{14}$N,  but   further  reactions   are  inhibited   by  the
$^{14}$N(p,$\gamma$) bottleneck.

The BSP08(GS) SSM \citep{bps08} -- which employs
values for $Z$ and the $^{14}$N(p,$\gamma$) S-factor given below -- predicts a
modest  CN-cycle  contribution  to   solar  energy  generation  of  0.8\%  but
substantial fluxes of neutrinos 
\begin{displaymath}
^{13}\mathrm{N} (\beta^+)^{13}\mathrm{C}~~E_\nu \lsim 1.199 
\mathrm{~MeV}~~\phi = (2.93^{+0.91}_{-0.82}) \times 10^8/\mathrm{cm^2s}
\end{displaymath} 
\begin{displaymath}
^{15}\mathrm{O} (\beta^+)^{15}\mathrm{N} ~~E_\nu \lsim 1.732
\mathrm{~MeV}~~\phi          =          (2.20^{+0.73}_{-0.63})          \times
10^8/\mathrm{cm^2s}. 
\end{displaymath}
Here uncertainties reflect conservative abundance  uncertainties  as defined
empirically in \citet{bs05}.
The first  reaction is part of the  path from $^{12}$C to  $^{14}$N, while the
latter  follows $^{14}$N(p,$\gamma$).   Thus neutrinos  from  $^{15}$O $\beta$
decay  are produced  in the  central core:  95\% of  the flux  comes  from the
CN-equilibrium region, described above.   About 30\% of the $^{13}$N neutrinos
come from outside  this region, primarily because of  the continued burning of
primordial  $^{12}$C: this  accounts for  the  somewhat higher  flux of  these
neutrinos.  There is  also a small but fascinating  contribution from $^{17}$F
$\beta$ decay,
\begin{displaymath}
^{17}\mathrm{F}(\beta^+)^{17}\mathrm{O}~~        E_\nu       \lsim       1.740
\mathrm{~MeV}~~\phi=(5.82\pm 3.04) \times 10^6/\mathrm{cm^2s} 
\end{displaymath}
a reaction  fed by (p,$\gamma$) on  primordial $^{16}$O: the  cycling time for
the second  branch of  the CNO  bi-cycle, for solar  core conditions,  is much
longer than the  solar age.  The flux of these neutrinos  appears too small to
allow  a  test   of  the  Sun's  primordial  oxygen   content  by  this  means
\citep{bahcallbook}.

The  SSM makes  several  reasonable assumptions,  including local  hydrostatic
equilibrium (the balancing of the gravitational force against the gas pressure
gradient), energy  generation by proton  burning, a homogeneous  zero-age Sun,
and boundary conditions  imposed by the known mass,  radius, and luminosity of
the  present Sun.   It assumes  no significant  mass loss  or  accretion.  The
homogeneity assumption  allows the primordial  core metalicity to be  fixed to
today's surface  abundances.  Corrections for  the effects of diffusion  of He
and the  heavy elements  over 4.57  b.y. of solar  evolution are  included, and
generally been helpful in improving  the agreement between SSM predictions and
parameters probed in helioseismology.

The assumption  of a homogeneous zero-age  Sun is based on  arguments that the
early pre-main-sequence Sun passed through a fully convective, highly luminous
Hayashi  phase,   homogenizing  the  Sun.   Yet,  as   recently  discussed  in
\citet{winnick}, whether this homogeneity persists
until the main sequence depends on  the Sun's metal accretion history.  In the
subsequent late pre-main-sequence phase (the Henyey phase), the Sun approaches
the  main sequence  by  establishing  and growing  a  radiative core.   Metals
accreted onto the Sun in or after this phase would not be mixed into the core.
Thus in  principle, if the accreted material  had a metal content  that is not
uniform in time,  differences between the surface and core  could arise in the
Henyey phase. \citet{winnick} have  discussed scenarios in  which such
accretion might produce  a convective zone enriched in  metals relative to the
radiative zone.  For many years one  motivation for models with such ``low Z''
cores was to  lower the $^8$B neutrino flux,  reducing the discrepancy between
the SSM and the results of the Davis experiment.

The SSM  assumes no such differentiation  occurs.  While this  assumption of a
homogeneous zero-age Sun may be correct, there are few observational checks on
proto-solar evolution.   But one possibility might be  the CN-cycle neutrinos.
The flux of  these neutrino should depend nearly linearly  on the initial core
abundance of C  and N.  If other uncertainties  affecting predictions of these
fluxes  could be brought  under control,  and if  these fluxes  were measured,
constraints on the core's primordial C and N might be obtained.

Solar surface  abundances are known, determined from  analyses of photospheric
atomic  and  molecular spectral  lines.   Traditionally  the associated  solar
atmosphere  modeling has  been done  in one  dimension, in  a time-independent
hydrostatic  analysis that incorporates  convection via  mixing-length theory.
But  much improved  3D  models of  the  solar atmosphere  have been  developed
recently  to treat  the radiation-hydrodynamics  and time  dependence  of this
problem.  This  approach is essentially  parameter-free and has been  shown to
accurately reproduce  average line  profiles, improve the  consistency between
different  line measurements  (e.g.,  among the  various  sources of  C and  O
lines), and bring the solar abundances  into better accord with other stars in
the solar neighborhood.  The  improved analysis, however, substantially lowers
the solar  metalicity from the  previous standard, Z=0.0169  \citep{gs98}, to
Z=0.0122  \citep{ags05},  and thus  alters  SSM  predictions. Hereafter  we
denote these as the GS and and AGS abundances, respectively. 

Solar models that use  the GS solar composition, the most up  to date of which
is the BPS08(GS) \citep{bps08} but including also the BP00 \citep{bpb00}, BP04
\citep{bp04}  and BS05(OP)  \citep{bsb05} models,  are in  excellent agreement
with those deduced from helioseismology.   But those computed with the revised
abundance are in 
much poorer  agreement, with  discrepancies exceeding 1\%  in the  region just
below  the   convective  zone  (R  $\sim  0.65-0.70   {\rm  R}_\odot$). 
Associated  properties of the SSM,  such as
the depth of the convective zone and the surface He abundance, are also now in
conflict  with helioseismology.   As extensively  discussed  in \citet{bsb_mc}
discrepancies   are   significantly   above   measurement  and   solar   model
uncertainties. 

The   reduced  core   opacity  also   lowers   the  SSM   prediction  of   the
temperature-dependent $^8$B  neutrino flux by about 20\%:  the predicted $^8$B
flux using  the GS abundances  and Opacity Project  \citep{opacity} opacities,
model BPS08(GS) is 5.95 $\times$ 10$^6$/cm$^2$s, 
 which drops  to 4.72  $\times$ 10$^6$/cm$^2$s when  AGS abundances  are used,
 model BPS08(AGS).  These results can be compared to the $^8$B neutrino flux
deduced from  the 391-day salt-phase  SNO data set  of 4.94 $\pm$  0.21 (stat)
$^{+0.38}_{-0.34}$    $\times$     10$^6$/cm$^2$s    \citep{sno_salt}.     The
Super-Kamiokande 
combined  unbinned (binned)  analysis using  the salt-phase  SNO data  finds a
best-fit flux  of 4.91  (4.86)$\times$ 10$^6$/cm$^2$ \citep{hosaka}.   Thus the
BPS08(GS) and BPS08(AGS) predictions are 1.2 and 0.95 times the experimental
central values  of the  combined analysis.  Both  results are  consistent with
experiment, given  current experimental (9.5\%) and  theoretical ($\sim$ 16\%)
uncertainties.

Finally,  we  describe a  speculative  scenario  to  illustrate why  a  direct
measurement of solar  core metalicity might be important  to our understanding
of solar system formation.  If one  were to attempt to construct a solar model
that reproduces both a sound-speed profile consistent with helioseismology and
the \citet{ags05}  photospheric abundances,  that model  would likely
have  a convective  zone that  is  {\it depleted}  in metals  relative to  the
radiative core,  not elevated as  in the low-Z  model familiar from  the solar
neutrino puzzle.   It is possible  to envision a  a scenario where  this could
happen --  one that connects the  chemistry of the Sun's  convective zone with
that of the  planets.  First, there is clear  evidence that solar-system metal
differentiation  occurred, associated  with  the formation  of the  metal-rich
gaseous giants Jupiter, Saturn,  Neptune, and Uranus.  The gaseous atmospheres
of  Jupiter and  Saturn are  believed to  have been  formed by  accretion onto
relatively small ($\sim$ 10 M$_\oplus$)  rocky/icy cores, at a time when solar
formation is nearly complete  and the bulk of the gas in  the nebular disk has
dissipated.  These atmospheres are  established over $\sim$ 1-10 million years
\citep{bod}.  It is thus plausible that the gaseous planets were formed after
the  sun had  developed  a radiative  core  and an  isolated convective  zone.
Second,  the amount  of  chemical  differentiation in  the  gaseous giants  is
suggestive: modelers estimate that Jupiter's  total metal content is between 8
and 39 M$_\oplus$, or Z $\sim$  2.5-12.3\%, while that of Saturn is between 13
and 28  M$_\oplus$, or Z  $\sim$ 13.7-29.4\% \citep{saumon}.  The  excess metal
contained in  all four  gaseous giants, $\sim$  40-90 M$_\oplus$  depending on
modeling uncertainties  \citep{guillot}, is comparable to  the apparent deficit
of metal in the convective zone  ($\sim$ 50 M$_\oplus$), were one to associate
the GS abundances with the radiative zone (formed from primordial gas) and the
AGS abundances  with the convective zone.  The  late-forming gaseous envelopes
of Jupiter and Saturn could account for up to 40 M$_\oplus$ of this excess.

Thus it is possible that some mechanism operating in a chemically altered disk
-- perhaps proto-planets scouring out metal-rich dust grains that have settled
to the disk  midplane -- might result both in metal  enrichment of the gaseous
giants and  a reservoir  of metal-depleted gas  in the  circum-planetary disk.
Could some  portion of that  gas later be  deposited on the Sun,  reducing the
effective  Z of  the  convective zone?   This  question has  been raised  once
before, by \citet{castro}, who then explored helioseismology in
a two-zone  model motivated by  this possibility.  While  we are
not advocating for  such a scenario, simple estimates,  including those above,
do seem to support its plausibility.  Numerical calculations indicate that the
time  scale  for the  Sun  to accrete  gas,  influenced  gravitationally by  a
Jupiter-mass body orbiting at a distance  $\sim$ 5 AU, is relatively short, on
the order of $  \sim 5 \times 10^5$ yr \citep{strom}.  It  has also been noted,
based on the needed planetesimal  deposition rate and the tidal radius ($\sim$
0.36 AU)  of a fully grown Jupiter,  that the planet would  have perturbed the
orbits of  about 2500  M$_\oplus$ of  gas, or 35\%  of the  mass of  the Sun's
present  convective  envelope  \citep{podolak}.   Thus  at least  some  of  the
conditions necessary for convective-zone dilution appear to be satisfied.

Though it is beyond the scope of  the present paper, it might be worthwhile to
pursue this  question further, assuming late-time  accretion of metal-depleted
gas (motivated by the gaseous giant metal reservoirs and their assumed time of
formation).   While  the  work by  \citet{castro}  is  a  first step  in  this
direction, if 
one takes the  accretion scenario seriously, then (depending  on the timing of
accretion  with respect to  the development  of the  radiative/convective zone
boundary) there should be a memory  of the accretion in the modern sun's upper
radiative zone --  a transition region between GS  interior abundances and AGS
surface abundances which depends on the volume and composition of the accreted
material.  Helioseismology would  thus become  a probe  of the  solar system's
late-stage  accretion  history.   This   history  would  be  linked  to  solar
development: in  the standard Hayashi-track description of  the proto-Sun, the
scenario would  only make sense if  the planet atmospheres were  formed in the
Henyey  phase  or  later.   However,  recent numerical  simulations  of  cloud
collapse and early solar evolution found that the convective envelope develops
earlier,  spans the  outer third  of the  proto-Sun by  radius,  and resembles
closely that of the modern  Sun \citep{wuchterl}.  This would extend the window
for dilution of  the convective zone to earlier  times.  Finally, the scenario
would predict chemical correlations between the planets and convective zone in
our sun and, perhaps, in other solar-like planetary systems.

The  various  threads  summarized  above, plus  additional  considerations  we
discuss  in this paper,  provide strong  motivation for  experiments measuring
CN-cycle neutrinos:
\begin{itemize}

\item A measurement of the CN  neutrino flux would provide an independent test
of solar metalicity, complementing photospheric determinations.

\item This measurement would test  the SSM postulate of a homogeneous zero-age
Sun, one of the assumptions important to helioseismology, the $^8$B neutrino flux,
and other SSM predictions that depend  on the metalicity of the Sun's interior
radiative zones.

\item It would  place constraints on metal accretion  that might have occurred
subsequent  to the Hayashi  phase, as  the pre-main-sequence  convective solar
core is established.

\item  The   current  solar  neutrino   program  has  helped   to  demonstrate
experimentally  that  the nuclear  astrophysics  foundations  of our  standard
theory of low-mass main-sequence  stellar evolution are valid.  Solar CN-cycle
neutrinos  provide our one  opportunity to  extend such  tests to  the nuclear
physics governing heavier main-sequence stars.

\item It is conceivable the a quantitative comparison of the Sun's surface and
interior metalicities might be important to more general problems of chemical
differentiation during solar-system formation.

\end{itemize}

\section{The Sun as a Calibrated Laboratory}

Independent  of questions  about  the Sun's  pre-main-sequence evolution,  one
recognizes that the Sun's inner core would have been mixed at the onset of the
main sequence due  to the initial out-of-equilibrium burning  of $^{12}$C.  It
has been  recognized for many years  that a measurement of  the CN-cycle solar
neutrino flux would, in principle, determine the metalicity of this core zone,
allowing  a comparison with  abundance determined  from the  solar atmosphere.
But in the past several years  new developments have occurred that now seem to
suggest such a measurement could be practical.  These include:
\begin{itemize}

\item  Accurate  calibrations  of  the  solar  core  temperature  by  SNO  and
Super-Kamiokande;

\item Tight constraints on the  oscillation parameters and matter effects that
will determine the flavor content of the CN and $^8$B neutrino fluxes;

\item  Recent  measurements of  the  controlling  reaction  of the  CN  cycle,
$^{14}$N(p,$\gamma$),  that  have significantly  reduced  the nuclear  physics
uncertainties affecting SSM predictions of CN-cycle fluxes; and

\item New ideas for high-counting  rate experiments that would be sensitive to
CN-cycle  neutrinos,  and from  which  reliable  terrestrial  fluxes could  be
extracted.

\end{itemize}
Our  analysis  uses  previous  SSM  work  in  which  the  logarithmic  partial
derivatives $\alpha(i,j)$  for each neutrino  flux $\phi_i$ are  evaluated for
the SSM input parameters $\beta_j$,
\begin{equation}
\alpha(i,j)  \equiv  {\partial   \ln{\left[  \phi_i/\phi_i(0)  \right]}  \over
\partial \ln{\left[ \beta_j / \beta_j(0)\right]}}
\end{equation}
where  $\phi_i(0)$  and  $\beta_j(0)$   denote  the  SSM  best  values.   This
information, in combination with  the assigned uncertainties in the $\beta_j$,
then  provides  an  estimate of  the  uncertainty  in  the SSM  prediction  of
$\phi_i$.   In particular,  crucial to  the current  analysis is  the  work of
\citet{bs05}, who evaluated the dependence on
the  mass  fractions  (measured  relative  to  hydrogen)  of  different  heavy
elements,
\begin{equation}
\beta_j =  \mathrm{mass~fraction~of~element~j \over mass~fraction~of~hydrogen}
\equiv X_j
\end{equation}
Having this information  not as a function of the  overall metalicity $Z$, but
as  a  function  of the  individual  abundances,  allows  us to  separate  the
``environmental'' effects  of the  metals in the  solar core from  the special
role of  primordial C and N as  catalysts for the CN  cycle.  By environmental
effects  we mean  the influence  of the  metals on  the opacity  and  thus the
ambient  core  temperature, which  controls  the  rates of  neutrino-producing
reactions of both the pp-chain and CN cycle.  Simply put, our strategy here is
to  use  the  temperature-dependent  $^8$B  neutrino  flux  to  calibrate  the
environmental  effects  of  the  metals  and of  other  SSM  parameters,  thus
isolating the  special CN-cycle  dependence on primordial  C+N.  We  find this
primordial abundance can  be expressed, with very little  residual solar model
uncertainty, in  terms of the measured  $^8$B neutrino flux  and nuclear cross
sections that have been determined in  the laboratory.  In fact, we argue that
the  resulting expression is  likely more  general than  the SSM  context from
which it is derived.

The  partial derivatives  allow one  to define  the power-law  dependencies of
neutrino fluxes, relative to the SSM best-value prediction $\phi_i(0)$
\begin{equation}
\phi_i   =   \phi_i(0)  \prod_{j=1}^N   \left[   {\beta_j  \over   \beta_j(0)}
\right]^{\alpha(i,j)}
\label{eq:prod}
\end{equation}
where the product extends over  $N$ SSM input parameters.  This expression can
be  used  to  evaluate  how  SSM  flux  predictions  will  vary,  relative  to
$\phi_i(0)$, as the  $\beta_j$ are varied.  Alternatively, the  process can be
inverted:  a flux  measurement  could in  principle  be used  to constrain  an
uncertain input parameter.

The baseline SSM calculation for our calculations is BPS08(AGS) \citep{bps08},
which uses the recently determined AGS
abundances for  the volatile  elements C, N,  O, Ne,  and Ar, rather  than the
previous  GS standard  composition. It  should be  noted that  AGS  includes a
downward revision by 0.05 dex of  the Si photospheric abundance compared to GS
and,  accordingly, a  similar  reduction in  the  meteoritic abundances.   The
partial derivatives needed in the present calculation are summarized in
Tables \ref{table:one} (solar model parameters and nuclear cross sections) and
\ref{table:two} (abundances).

The SSM  estimate  of  uncertainties  in the  various  solar
neutrino fluxes 
$\phi_i$  can  be  obtained  by  folding  the  partial  derivatives  with  the
uncertainties in the underlying $\beta_j$.  In particular, it is convenient to
decompose Eq.  \ref{eq:prod} into its  dependence on solar  parameters, non-CN
metals, nuclear S-factors, and the primordial C and N abundances,
\begin{eqnarray}
\phi_i =  \phi_i^{SSM} \times~~~~~~~~~~~~~~~~~~~~~~~~~~~~~~~~~~~~~~~ \nonumber
\\ \left(  \prod_{j \in \mathrm{\{Solar\}}} \left[  {\beta_j \over \beta_j(0)}
\right]^{\alpha(i,j)}  \prod_{j  \in   \mathrm{\{Metals  \neq  C,N\}}}  \left[
{\beta_j \over  \beta_j(0)} \right]^{\alpha(i,j)} \right)  \nonumber \\ \times
\prod_{j   \in  \mathrm{\{Nuclear\}}}   \left[   {\beta_j  \over   \beta_j(0)}
\right]^{\alpha(i,j)}  \prod_{j \in  \mathrm{\{C,N\}}}  \left[ {\beta_j  \over
\beta_j(0)} \right]^{\alpha(i,j)} .~~~~~~~
\label{eq:prod2}
\end{eqnarray}

The  two  terms  within  the  brackets will  be  designated  ``environmental''
uncertainties --  SSM solar and abundance parameters  that primarily influence
neutrino flux predictions through changes they induce in the core temperature.
These are, respectively, the uncertainties in the photon luminosity $L_\odot$,
the  mean  radiative opacity,  the  solar age,  and  calculated  He and  metal
duffusion; and the fractional abundances of O, Ne, Mg, Si, S, Ar, and Fe.  The
estimated  1$\sigma$ fractional  uncertainties for  the solar  parameters have
been previously evaluated and are listed in Table \ref{table:three}.

The heavy elements abundances in
BPS08(AGS) are taken from  the meteoritic abundances where available (Mg,
Si, S,  and Fe) and otherwise  from photospheric abundances  (for the volatile
elements   C,   N,  O,   Ne,   Ar).   As   mentioned  before,   the   assigned
historical/conservative 1$\sigma$ fractional 
uncertainties  shown in  Table  \ref{table:four} were  defined empirically  in
\citet{bs05} by
\begin{equation}
{\Delta \beta_i \over  \beta_i} = \left| {\mathrm{Abundance}_i^{\mathrm{GS}} -
\mathrm{Abundance}_i^{\mathrm{AGS}}  \over (\mathrm{Abundance}_i^{\mathrm{GS}}
+ \mathrm{Abundance}_i^{\mathrm{AGS}})/2)} \right|.
\end{equation}
This definition generates the uncertainties shown in Table \ref{table:four}.

The next term  contains the effects of nuclear  cross section uncertainties on
flux predictions.  The $\beta_j$ are  the S-factors for p+p ($S_{11}$), $^3$He
+  $^3$He ($S_{33}$),  $^3$He+$^4$He ($S_{34}$),  p +  $^7$Be ($S_{17}$),  e +
$^7$Be ($S_{e7}$),  and p +  $^{14}$N ($S_{114}$).  Their  estimated 1$\sigma$
fractional  uncertainties, which  we discuss  below, are  also shown  in Table
\ref{table:three}.

The last  term is the contribution of  the primordial C and  N abundances.  As
Table   \ref{table:two}  shows,  pp-chain   neutrino  fluxes   are  relatively
insensitive to variations  in these abundances, as the  heavier nuclei like Fe
have a more important influence on the core opacity.  But the expected, nearly
linear  response  of  the  $^{13}$N  and $^{15}$O  neutrino  fluxes  to  these
abundances is apparent.   These are the abundances we  would like to constrain
by a  future measurement of the  $^{13}$N and $^{15}$O  solar neutrino fluxes.
Such  a measurement  begins to  be of  interest if  these abundances  could be
determined with an accuracy significantly better than 30\%.  Note that the C/N
abundance term in Eq. (\ref{eq:prod2}) for $\phi(^{13}\mathrm{N})$
\begin{eqnarray}
 \prod_{j   \in   \mathrm{\{C,N\}}}    \left[   {\beta_j   \over   \beta_j(0)}
 \right]^{\alpha(i,j)}=  ~~~~~~~~~~~~~\nonumber \\  \left( {X(^{12}\mathrm{C})
 \over  X(^{12}\mathrm{C})_{SSM}}  \right)^{0.874} \left(  {X(^{14}\mathrm{N})
 \over X(^{14}\mathrm{N})_{SSM}} \right)^{0.142}
\end{eqnarray}
is not quite linear in the overall C+N abundance.  An overall scaling of
primordial C  and N, $X/X_{SSM} \rightarrow$ 1+$\delta$  yields the dependence
$(1+\delta)^{1.016}$, so 2\%  steeper than linear.  In addition  to the direct
dependence  of the  CN-cycle on  C and  N, increasing  the C+N  abundance also
increases the opacity and core temperature, to which the CN neutrino flux also
responds.  

Were  one  to vary  the  11  SSM  parameters designated  as  ``environmental''
according to their assigned uncertainties (taking them to be uncorrelated), an
7.5\% SSM  net uncertainty in $\phi(^{13}\mathrm{N})$ would  be obtained.  But
we can do better than this by exploiting SNO and Super-Kamiokande measurements
of the $^8$B  neutrino flux, a ``thermometer'' that is  even more sensitive to
the solar core environment than the CN neutrinos.  Below we discuss the use of
SNO in  this way --  the arguments are  simpler for this detector,  because it
provides a measurement  of the $^8$B neutrino flux  independent of flavor.  In
the next section we use Super-Kamiokande, the case of most interest because it
exploits the  same reaction, elastic  scattering, as the  proposed CN-neutrino
detectors, allowing some common errors  to cancel: the SNO results then become
crucial input, helping to constrain the effects of neutrino oscillations.

\begin{figure*}
\includegraphics[width=16cm]{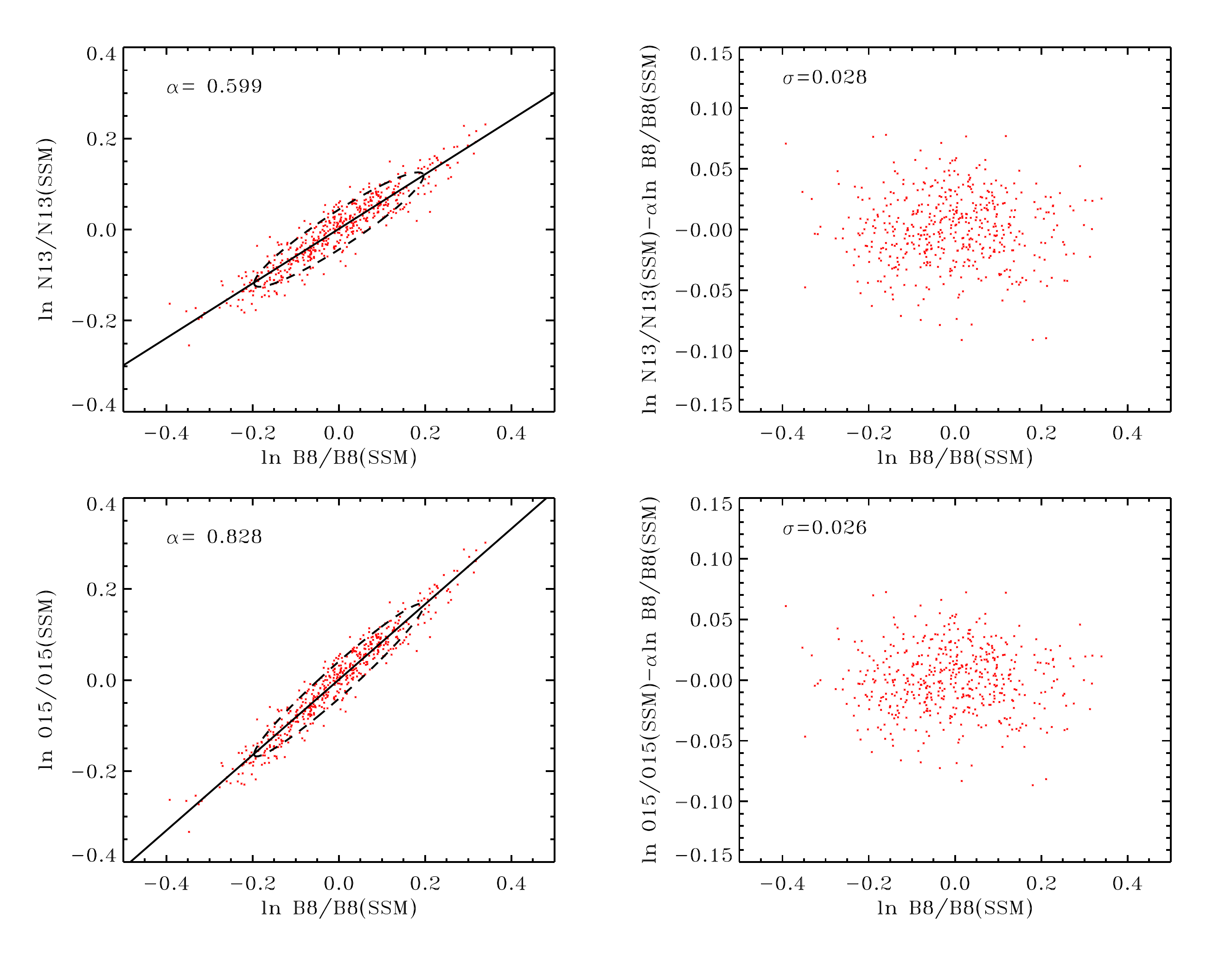}
\caption{Results  from  a   Monte  Carlo  simulation  of  SSM   where  the  11
  environmental parameters  (see text) have  been varied. The two  left panels
  show  the correlations between  the $^8{\rm  B}$ flux  and the  two CN-cycle
  neutrino  fluxes  $^{13}{\rm N}$  and  $^{15}{\rm  O}$.  The slopes  of  the
  correlations are  given in  the plots, together  with the  68.3\% confidence
  level contours.  On the  right side  panels we show  the residuals  from the
  fits,  2.8\% and  2.6\% for  the $^{13}{\rm  N}$ and  $^{15}{\rm  O}$ fluxes
  respectively,  that  determine  the  residual environmental  uncertainty  in
  Eqs.~(\ref{eq:prod4},\ref{eq:prod4b}) respectively.  \label{fig:correl}} 
\end{figure*}

One can express $\phi(^{13}\mathrm{N})$
uncertainties in terms of $\phi(^8\mathrm{B})$, while minimizing the residual
solar environmental error,  i.e. by minimizing the contribution  of the factor
in parenthesis in the expression
\begin{eqnarray}
&&{\phi     (^{13}\mathrm{N})     \over     \phi^{SSM}(^{13}\mathrm{N})}     =
\left[{\phi(^{8}\mathrm{B})                                               \over
    \phi^{SSM}(^{8}\mathrm{B})}\right]^{K_{(13,8)}} 
\nonumber \\ &\times& \left(  \prod_{j \in \mathrm{\{Solar\}}} \left[ {\beta_j
\over  \beta_j(0)}  \right]^{\gamma_{j}}  \prod_{j \in  \mathrm{\{Metals  \neq
C,N\}}}  \left[   {\beta_j  \over  \beta_j(0)}   \right]^{\gamma_{j}}  \right)
\nonumber \\ &\times& \prod_{j \in \mathrm{\{Nuclear\}}} \left[ {\beta_j \over
\beta_j(0)}   \right]^{\gamma_{j}}  \prod_{j   \in   \mathrm{\{C,N\}}}  \left[
{\beta_j \over \beta_j(0)} \right]^{\gamma_{j}}
\label{eq:prod3}
\end{eqnarray}
where
\begin{equation}
\gamma_{j}       \equiv      \alpha(^{13}\mathrm{N},j)       -      K_{(13,8)}
\alpha(^8\mathrm{B},j)
\end{equation}
by a suitable choice of the constant $K_{(13,8)}$. 
Using   the   SSM   logarithmic   derivatives  $\alpha(i,j)$   and   parameter
uncertainties         $\Delta        \beta_j/\beta_j$         of        Tables
\ref{table:one}-\ref{table:four},  we find $K_{(13,8)}=  0.608$. To  check the
consistency of this  procedure, we have performed a  Monte Carlo simulation of
solar  models   where  the  11   environmental  input  quantities   have  been
varied   simultaneously.    We   find   a  tight   correlation   between   the
$\phi(^{13}{\rm N})$
and  $\phi(^{8}{\rm  B})$  fluxes.  In Figure~\ref{fig:correl}  we  show  this
correlation on the  top-left panel and the linear fit to  the data, from which
we find $K_{(13,8)}= 0.599$, the
value we adopt  for this paper, very close to that  derived from the power-law
exponents. The top-right panel shows the residuals of the fit and its standard
deviation $\sigma=2.8\%$.  We note here that the bulk of the dispersion in the
$\phi(^{8}{\rm B})-\phi(^{13}{\rm  N})$ correlation is due  to the uncertainty
in the 
diffusion  rate.   This  can  be  understood  as  follows.  All  environmental
quantities 
affect  these  neutrino fluxes  by  modifying  the  temperature in  the  solar
core. Diffusion has,  however, the additional effect of  increasing the number
of CN nuclei  in the core, leading to a  directly proportional increase in
$\phi(^{13}{\rm 
N})$   but   not   in   $\phi(^{8}{\rm    B})$.    This   can   be   seen   in
Tables~\ref{table:one} and 
\ref{table:two},   where  all  the   power-law  exponents   for  environmental
quantities are larger for $\phi(^{8}{\rm B})$ than for $\phi(^{13}{\rm N})$ (a
natural    consequence    of   the    larger    temperature   dependence    of
$\phi(^{8}{\rm   B})$).   The   only   exception  is   diffusion,   on   which
$\phi(^{13}{\rm N})$ 
shows  a stronger  dependence than  $\phi(^{8}{\rm  B})$. Were  we to  exclude
diffusion as a source of 
uncertainty, the dispersion in the $\phi(^{8}{\rm B})-\phi(^{13}{\rm N})$
correlation would only be $\sim 0.5\%$.

Expressing $\phi(^{13}{\rm N})$ in the form of Eq.~(\ref{eq:prod3})
has two advantages.  First, as we detail below, it 
reduces the  overall theoretical uncertainty  in the relationship  between the
primordial C and N abundances and the $^{13}$N neutrino flux.  Second, this 
relationship should  be more  general than  the SSM context  from which  it is
derived:  the correlation  between the  various neutrino  fluxes  $\phi_i$ and
$T_c$ has been demonstrated to hold  even when SSM parameters have been varied
far outside their accepted SSM uncertainties.  

A   simple  way   to  fix   the  first   term  on   the  right-hand   side  of
Eq.  (\ref{eq:prod3}) is  by  using the  SNO  measurement of  the total  $^8$B
neutrino  flux, thereby  eliminating many  SSM and  LMA  oscillation parameter
uncertainties in  terms of  a measured quantity.   As the SNO  statistical and
systematic errors  combined in quadrature  give an uncertainty 9.4\%,  the net
uncertainty  in  this  term   is  then  5.6\%.   The  remaining  environmental
uncertainty is  encoded in  the bracketed terms  in Eq.  (\ref{eq:prod3}), the
deviations in the dependence of the  $^{13}$N and $^8$B neutrino fluxes from a
naive $T_c$ power law.  From our  Monte Carlo simulation we find this residual
uncertainty is 2.8\%, and thus
quite small in comparison to the SNO uncertainty: the total environmental
uncertainty  is $\sim$  6.3\%.  The  use of  the SNO  result to  constrain the
environmental  uncertainty  thus reduces  this  uncertainty,  relative to  the
result obtain  previously by direct  variation of SSM input  parameters within
their assigned SSM uncertainties.

The remaining uncertainty arising in the evaluation of Eq. (\ref{eq:prod3}) is
the nuclear factor which, given that the expression involves both the $^{14}$N
and $^8$B fluxes, depends on a combination of pp-chain and CN-cycle S-factors.
One finds from Tables \ref{table:one}- \ref{table:four} that the uncertainties
are dominated by $S_{17}$, $S_{34}$,  which controls the pp-chain branching to
the ppII and ppIII cycles, and $S_{114}$.  One of the reasons that the CN
neutrino fluxes are potentially a quantitative probe of the Sun's primordial C
and  N  are  recent  improvements  particularly  in  determinations  of  the
last two $S$-factors.  

The traditional SSM value for $S_{34}$ is based on the 1998 evaluation of 
\citet{adelberger}, 0.53 $\pm$ 0.05 keV b.  The relatively large error
bar  on  the recommended  value  reflected  apparent systematic  disagreements
between  experiments detecting prompt  $\gamma$ rays  and counting  the $^7$Be
activity.  Since this evaluation new, high-statics measurements have been made
by  a Weizmann  Institute  group (\citealt{singh},  an
activity  measurement),  by the  LUNA  collaboration (\citealt{confortola},  a
combination of prompt $\gamma$ and activity measurements, 
as well as \citealt{gyurky}, an activity measurement), and
by  the  Seattle   group  (\citealt{brown},  an  activity
measurement).   If these  data are  extrapolated  to threshold  with the  same
theoretical fitting function,  one finds values of S(0)  of 0.546 $\pm$ 0.020,
0.560  $\pm$  0.017,  0.545  $\pm$   0.017,  and  0.595  $\pm$  0.018  keV  b,
respectively.  The spread  of these  results is  somewhat larger
than is expected  on the basis of the uncertainties.   They have been combined
by Snover (private communication) in a way that takes into account this spread
by 
inflating  the errors  according  to the  associated  chi-squares, while  also
accounting  for possible  correlations  between the  LUNA  activity and  total
results  \citep{gyurky,confortola}.  The  result is  0.564 $\pm$  0.020  keV b.
There is an additional theoretical uncertainty associated with the theoretical
fitting function that is used to extrapolate these data to threshold to obtain
S(0).  As  discussed in \citet{brown}, generally  this variation is  at the few
percent  level, though  it can  reach higher  values if  fits are  required to
reproduce higher-energy data.  The procedure  we followed is based on the LUNA
group's  work on its  activation data  \citep{gyurky}, where  three theoretical
models  (\citealt{kajino,csoto}  and  \citealt{descouvemount})  were  used  to
derived  an extrapolated zero-energy  result.  The  resulting best  values for
$S(0)$ have a range of $\pm$ .019  keV b, or $\pm$ 3.4\%.  Thus we adopt 3.4\%
as  a theoretical extrapolation  uncertainty, which  we combine  in quadrature
with the Snover recommendation to obtain  a final result of 0.564 $\pm$ 0.028.
This corresponds to a 4.9\% uncertainty, which can be compared to the 9.4\%
uncertainty recommended in the \citet{adelberger} evaluation.

The most important  nuclear physics uncertainty in the  analysis is $S(0)$ for
$^{14}$N(p,$\gamma$), a reaction  that has been the subject  of recent intense
study.    The  $S_{114}$   value  and   the  8.4\%   uncertainty   adopted  in
\citet{bsb_mc}, 1.69 $\pm$ 0.14 keV b, was 
obtained from combining  the LUNA results of \citet{formicola},  1.7 $\pm$ 0.2
keV b (determined from data taken at or above 
center-of-mass energies of 140 keV),  with the TUNL results of \citet{runkle},
1.68 $\pm$ 0.09 (stat) $\pm$ 0.16 (sys) keV b.  
Subsequently a  series of  measurements have  been done at  LUNA in  which the
cross  section  was measured  to  center-of-mass energies  as  low  as 70  keV
\citep{imbriani,lemut,bemmerer,traut}.   In particular,  \citet{imbriani} give
$S(0)$ = 1.61 $\pm$ 0.08 keV b, corresponding to a 
5\% error, based on data  obtained for center-of-mass energies between 119 and
367 keV.  We  use the Imbriani value for $S(0)$ in  this paper, a conservative
choice given that this fit was  made prior to the extension of measurements to
70 keV.   Furthermore, work is underway  on the energy range  above the lowest
resonance, $\sim$ 300-400 keV, a  region which limits the interference pattern
analysis, and on improved  r-matrix analyses (Wiescher, private communication)
which take into account all reaction channels.  Thus we expect a more definite
analysis of $S(0)$ and of  its experimental and theoretical uncertainties will
be available soon. 

All of this can then be summarized in the theoretical relationship:
\begin{eqnarray}
&&{\phi     (^{13}\mathrm{N})     \over     \phi^{SSM}(^{13}\mathrm{N})}     =
\left[{\phi^{SNO}(^{8}\mathrm{B})                                         \over
\phi^{SSM}(^{8}\mathrm{B})}\right]^{0.599} \nonumber \\  &\times& \left[ 1 \pm
2.8\%   (\mathrm{resid.~environ.})  \pm   5.0\%)   (\mathrm{nuclear})  \right]
\nonumber      \\      &\times&      \left(     {X(^{12}\mathrm{C})      \over
X(^{12}\mathrm{C})_{SSM}}  \right)^{0.858}  \left(  {X(^{14}\mathrm{N})  \over
X(^{14}\mathrm{N})_{SSM}} \right)^{0.141}.
\label{eq:prod4}
\end{eqnarray}
The first term,  given the SNO total flux combined error  of $\sim$ 9.4\%, has
an uncertainty of 6.4\%.  Thus one  consequence of the recent work on pp-chain
and CN-cycle S-factors  is the reduction of the  nuclear physics uncertainties
to the  level of  the SNO measurements.   The overall uncertainty,  adding the
SNO,  residual  environmental, and  nuclear  uncertainties  in quadrature,  is
8.6\%,  small compared  to the  conservative uncertainty  assigned to  the AGS
abundances of $\sim$ 30\%.

Under  an overall  scaling  of  primordial C  and  N, $X/X_{SSM}  \rightarrow$
1+$\delta$, the quantity being constrained responds as
\begin{equation}
(1+\delta)^{0.999} \sim 1 + \delta
\label{eq:delta2}
\end{equation}
Because  we  have removed  the  ``environmental''  effects  of all  metals  in
Eq.   (\ref{eq:prod4}),   we   find   the  expected,   nearly   exact   linear
proportionality  between the primordial  metals C  and N  and the  CN neutrino
flux.

The arguments can be repeated for the $^{15}$O flux, the more interesting case
experimentally because of its higher endpoint energy.  We find
\begin{eqnarray}
&&{\phi     (^{15}\mathrm{O})     \over     \phi^{SSM}(^{15}\mathrm{O})}     =
\left[{\phi(^{8}\mathrm{B})  \over  \phi^{SSM}(^{8}\mathrm{B})}\right]^{0.828}
\nonumber \\ &\times& \left(  \prod_{j \in \mathrm{\{Solar\}}} \left[ {\beta_j
\over  \beta_j(0)}  \right]^{\gamma_{j}}  \prod_{j \in  \mathrm{\{Metals  \neq
C,N\}}}  \left[   {\beta_j  \over  \beta_j(0)}   \right]^{\gamma_{j}}  \right)
\nonumber \\ &\times& \prod_{j \in \mathrm{\{Nuclear\}}} \left[ {\beta_j \over
\beta_j(0)}   \right]^{\gamma_{j}}  \prod_{j   \in   \mathrm{\{C,N\}}}  \left[
{\beta_j \over \beta_j(0)} \right]^{\gamma_{j}}
\label{eq:prod3a}
\end{eqnarray}
where
\begin{equation}
\gamma_{j} \equiv \alpha(^{15}\mathrm{O},j) - 0.828 \alpha(^8\mathrm{B},j).
\end{equation}
The  lower two panels  in Figure~\ref{fig:correl}  show the  tight correlation
between the  $\phi (^{15}\mathrm{O})$  and $\phi (^{8}\mathrm{B})$  fluxes and
the residuals of the linear fit. 

Evaluating associated parameter uncertainties as before, one finds
\begin{eqnarray}
&&{\phi     (^{15}\mathrm{O})     \over     \phi^{SSM}(^{15}\mathrm{O})}     =
\left[{\phi^{SNO}(^{8}\mathrm{B})                                         \over
\phi^{SSM}(^{8}\mathrm{B})}\right]^{0.828} \nonumber \\  &\times& \left[ 1 \pm
2.6\%   (\mathrm{resid.~environ.})   \pm   7.1\%  (\mathrm{nuclear})   \right]
\nonumber      \\      &\times&      \left(     {X(^{12}\mathrm{C})      \over
X(^{12}\mathrm{C})_{SSM}}  \right)^{0.805}  \left(  {X(^{14}\mathrm{N})  \over
X(^{14}\mathrm{N})_{SSM}} \right)^{0.199}
\label{eq:prod4b}
\end{eqnarray}
The larger  environmental parameter, 0.828,  is expected because  the $^{15}$O
neutrino  flux  has  a  somewhat   steeper  dependence  on  the  average  core
temperature than the $^{13}$N flux.  This increases the errors associated with
SNO  uncertainties (7.8\%)  and  nuclear cross  sections  (the uncertainty  in
$S_{34}$ and other pp-chain cross sections determines the quality of our $^8$B
neutrino thermometer).   Thus the overall uncertainty  in this ``theoretical''
relation  between the $^{15}$O  neutrino flux  and the  core C/N  abundance is
10.8\%.   As  before, if  one  considers  a scaling  of  primordial  C and  N,
$X/X_{SSM} \rightarrow$ 1+$\delta$, the last term becomes
\begin{equation}
(1+\delta)^{1.004} \sim 1 + \delta
\label{eq:delta3}
\end{equation}
showing the  linear relationship  between the $^{15}$O  neutrino flux  and the
primordial C+N abundance.

Now we discuss a somewhat  more detailed (and improved) analysis that exploits
similarities between  Super-Kamiokande and  future CN neutrino  detectors, and
allows us  to fold in  what we have  learned about neutrino  oscillations from
terrestrial experiments like KamLAND.

\section{The Analysis for Elastic Scattering and Neutrino Oscillations}

Equations (\ref{eq:prod3}) and  (\ref{eq:prod3a}) give relationships among the
primordial  core  C  and  N  abundances, other  SSM  uncertainities,  and  the
instantaneously produced CN-cycle and $^8$B neutrino fluxes.  These equations,
attractive because  of their simplicity, are somewhat  idealized, because they
do not address  how the left-hand sides of these  equations will be determined
experimentally.  For example, neutrino  oscillations during the transit to the
earth will  alter the  flavors of these  neutrino in an  energy-dependent way,
influencing detector  responses.  Fluxes  determined in most  experiments will
have  to be corrected  for such  effects, including  the uncertainties  in the
neutrino  mass  difference  $\delta   m_{12}^2$  and  mixing  angle  $\sin^2{2
\theta_{12}}$.  In  the discussion  of the previous  section, we  avoided this
issue  in  evaluating  the   right-hand  sides  of  Eq.  (\ref{eq:prod3})  and
(\ref{eq:prod3a}) by  employing the  SNO result for  the total  $^8$B neutrino
flux.  But the  first results on the CN-cycle fluxes,  needed on the left-hand
sides of Eqs. (\ref{eq:prod3}) and  (\ref{eq:prod3a}), are most likely to come
from  $\nu$-e   inelastic  scattering  experiments,   where  $\sigma(\nu_\mu)/
\sigma(\nu_e) \sim$ 0.15.  Thus to  derive the instantaneous (solar) values of
these fluxes, one would have to  correct the detector response for the effects
of flavor  mixing.  Oscillations  are one of  several uncertainties  that will
produce correlated responses in both $^8$B and CN neutrino detectors.  Thus we
need an  analysis that accounts for  such correlations: it  is advantageous to
develop  this analysis,  as one  can then  make use  of  another statistically
powerful experiment (Super-Kamiokande) while supplementing SNO data with other
constraints on flavor mixing (e.g., KamLAND).

Below  we compare Super-Kamiokande  and Borexino/SNO+  rates, which  exploit a
common  detection mechanism, $\nu_x$-e  elastic scattering,  making correlated
errors  easier to identify.   Super-Kamiokande is  potentially the  best solar
thermometer because of its statistical precision.

In this approach we express the total $^8$B flux (the instantaneous solar flux
needed in Eqs. (\ref{eq:prod3}) and (\ref{eq:prod3a})) as
\begin{eqnarray}
{\phi(^8\mathrm{B})   \over  \phi^{SSM}(^8\mathrm{B})}&=&  {\phi(^8\mathrm{B})
\langle  \sigma^{SK}(^8\mathrm{B}, \delta m_{12}^2,\theta_{12})  \rangle \over
\phi^{SSM}(^8\mathrm{B})      \langle     \sigma^{SK}(^8\mathrm{B},     \delta
m_{12}^2,\theta_{12})      \rangle      }      \nonumber      \\      &\equiv&
{R^{SK}_{exp}(^8\mathrm{B})     \over     R^{SK}_{cal}(^8\mathrm{B},    \delta
m_{12}^2,\theta_{12})}
\label{eq:fluxratio}
\end{eqnarray}
Here $\langle  \sigma^{SK} \rangle$ is  an effective cross section  that takes
into  account  all of  the  neutrino  flavor  and detector  response  (trigger
efficiencies,  resolution,  cross  section  uncertainties, etc.)  issues  that
determine  the   relationship  between  a  measured  detector   rate  and  the
instantaneous  solar flux.   The numerator  of  the ratio  on the  right is  a
directly   measured  experimental   quantity:  the   Super-Kamiokande  elastic
scattering rate for producing  recoil electrons with apparent energies between
5.0  and  20 MeV,  per  target  electron per  second.   The  denominator is  a
conversion  factor  that  relates   the  instantaneous  $\nu_e$  flux  to  the
experimental  rate:  the  cross  section  for  $\nu_x-e$  elastic  scattering,
averaged  over   a  normalized  $^8$B  spectrum,  defined   for  the  specific
experimental  conditions of  Super-Kamiokande,  and including  the effects  of
flavor mixing.   This conversion factor is essentially  a laboratory quantity:
it can be calculated from  laboratory measurements of detector properties, the
$\beta$ decay  spectrum, the underlying neutrino-electron  cross sections, and
most  critically, the  parameters governing  oscillations.  We  describe these
factors below.

The  experimental   rate  comes  from   the  1496  days  of   measurements  of
Super-Kamiokande I \citep{hosaka}.  From the SK I rate/kiloton/year
\begin{equation}
520.1        \pm         5.3        \mathrm{(stat)}        ~{}^{+18.2}_{-16.6}
    \mathrm{(sys)}~\mathrm{kton^{-1}~y^{-1}}.
\label{eq:rate1}
\end{equation}
we find $R^{SK}_{exp}(^8\mathrm{B})$,
\begin{equation}
4.935    \pm    0.05   \mathrm{(stat)}~{}^{+0.17}_{-0.16}\mathrm{(sys)}~\times
  10^{-38}~\mathrm{{electron}^{-1} s^{-1}}
\end{equation}
(or  $\sim$ 0.049  Solar Neutrino  Units, or  SNUs).  The  dominant systematic
error  includes  estimates  for  the  energy  scale  and  resolution,  trigger
efficiency, reduction, spallation dead time,  the gamma ray cut, vertex shift,
background  shape  for  signal  reduction, angular  resolution,  and  lifetime
uncertainties.  The  combined statistical and  systematic error is  $\sim \pm$
3.6\%.

To evaluate the denominator in  Eq. (\ref{eq:fluxratio}) we need the suitably
averaged cross section, defined for the window used by the SK I collaboration,
\begin{eqnarray}
 \langle \sigma^{SK}(^8\mathrm{B},  \delta m_{12}^2,\theta_{12}) \rangle= \int
d  E_\nu  \phi_{norm}^{^8\mathrm{B}}(E_\nu)~~~~~~~~~~~  \nonumber &&\\  \times
\left[       P_{\nu_e}(E_\nu,\delta       m^2_{12},\sin^2{2      \theta_{12}})
\int_{T=0}^{T^{max}(E_\nu)}   dT    ~\sigma_{\nu_e}^{es}(T)   \right.   ~~~~&&
\nonumber    \\   \left.    +    ~P_{\nu_\mu}(E_\nu,\delta   m^2_{12},\sin^2{2
\theta_{12}}) \int_{T=0}^{T^{max}(E_\nu)} dT ~\sigma_{\nu_\mu}^{es}(T) \right]
~~~&&   \nonumber  \\  \times   \int_{5.0~\mathrm{MeV}}^{20.0~\mathrm{MeV}}  d
\epsilon_a                                     f_{\mathrm{trigger}}(\epsilon_a)
\rho(\epsilon_a,\epsilon_t=T+m_e)~~~~~~
\label{eq:theorypart}
\end{eqnarray}
where  $\phi_{norm}^{^8\mathrm{B}}(E_\nu)$ is  the  normalized $^8$B  neutrino
spectrum.   Equation  (\ref{eq:theorypart})  involves  an  integral  over  the
product    of   this   spectrum    and   the    energy-dependent   oscillation
probabilities.  ($P_{\nu_e}+P_{\nu_\mu}$=1, assuming oscillations  into active
flavors.  $P_{\nu_\mu}$ can be defined as the oscillation probability to heavy
flavors, if  the effects  of three flavors  are considered.)  A  given $E_\nu$
fixes the range of kinetic energies  $T$ of the scattered electron, over which
an integration  is done; in the  laboratory frame $T^{max}  = 2 E_\nu^2/(m_e+2
E_\nu)$.   The  integrand  includes  the  elastic  scattering  cross  sections
$\sigma^{es}(T)$   for   electron   and   heavy-flavor   neutrinos   and   the
Super-Kamiokande  resolution   function  $\rho(\epsilon_a,\epsilon_t)$,  where
$\epsilon_t=T+m_e$  is the true  total electron  energy while  $\epsilon_a$ is
apparent energy, as deduced from the number of phototube hits in the detector.
Finally,   an  integral   must  be   done  over   the  window   used   by  the
experimentalists, apparent electron energies $\epsilon_a$ between 5.0 and 20.0
MeV.  The  deduced counting  rate includes the  triggering probability  that a
event of apparent energy $\epsilon_a$ will be recorded in the detector.
        
As the  uncertainties associated  with triggering efficiencies,  energy scale,
and resolution are already  incorporated in the deduced Super-Kamiokande event
rate (Eq.~\ref{eq:rate1}),  we are free to use  best-value functions in such
an  analysis.   Here  we employ  simple  fits  to  the measurements  given  in
\citet{hosaka}, a Gaussian resolution function
\begin{equation}
\rho(\epsilon_a,\epsilon_t)={1  \over  \sqrt{2  \pi} \sigma(\epsilon_t)}  \exp
\left[ -{(\epsilon_t-\epsilon_a)^2 \over 2 \sigma(\epsilon_t)^2} \right]
\label{eq:resolution}
\end{equation}
where $\sigma(\epsilon_t) \sim 0.326  \epsilon_t^{0.642}$, or about 14\% at 10
Mev; and a relatively sharp trigger efficiency
\begin{equation}
f_{\mathrm{trigger}}(\epsilon_a)   =    {1   \over   2}   +    {1   \over   2}
\tanh^5{\left[{\epsilon_a-\bar{\epsilon}_a \over \sigma'}\right]}
\end{equation}
where $\bar{\epsilon}_a \sim$ 3.6 MeV and $\sigma' \sim$ 0.172 MeV.

The   expression  for   the   CN-cycle  neutrino   response   is  similar   to
Eq. (\ref{eq:fluxratio})
\begin{eqnarray}
{\phi(^{15}\mathrm{O})       \over       \phi^{SSM}(^{15}\mathrm{O})}      &=&
{R_{exp}^{B/S}(^{15}\mathrm{O})   \over   \phi^{SSM}(^{15}\mathrm{O})  \langle
\sigma^{B/S}(^{15}\mathrm{O}, \delta  m_{12}^2,\theta_{12}) \rangle} \nonumber
\\      &\equiv&     {      ~R^{B/S}_{exp}(\mathrm{^{15}\mathrm{O}})     \over
R^{B/S}_{cal}(^{15}\mathrm{O}, \delta  m_{12}^2,\theta_{12})} \nonumber \\ &=&
{R_{exp}^{B/S}(\mathrm{CN})/(1             +\alpha(0.8,1.3))             \over
R^{B/S}_{cal}(^{15}\mathrm{O}, \delta m_{12}^2,\theta_{12})) }
\label{eq:fluxratio2}
\end{eqnarray}
Here the experimental rate for $^{15}$O neutrinos has been written in terms of
the  total  CN-neutrino  rate  $R^{B/S}_{exp}(\mathrm{CN})$ by  introducing  a
correction   factor    $\alpha$   discussed   below.     No   measurement   of
$R^{B/S}_{exp}(\mathrm{CN})$   currently   exists,   of  course.    But   such
measurements could be made in  Borexino or SNO+, existing or planned detectors
that  will  use large  volumes  of  organic  scintillator, placed  quite  deep
underground (we discuss these detectors in the concluding section).  A
window  for the  apparent  kinetic energy  $T$  of the  scattered electron  of
0.8-1.3 MeV has been discussed by the Borexino group.  As the $^7$Be 0.866 MeV
line  corresponds  to  $T^{max}  \sim$  0.668 MeV,  this  window  would  limit
contamination from $^7$Be neutrino recoil electrons.  As Borexino has achieved
a resolution of $\sim$  8.7\% at $T$ = 751 MeV and  $\sim$ 9.1\% at 0.825 MeV,
we   take   (for  simulation   purposes)   a   Gaussian  resolution   function
(Eq.~\ref{eq:resolution}) with
\begin{equation}
\sigma(T) \sim 0.08 \mathrm{MeV} \sqrt{T/\mathrm{MeV}}
\end{equation}
We  also  adopt  a  nominal  step-function  trigger,  $f_{\mathrm{trigger}}  =
\theta(T-0.25 {\mathrm{MeV}})$, though the trigger does not influence rates in
the high-energy window of interest for CN neutrinos.

The factor $\alpha(0.8,1.3) \sim$ 0.120  is the ratio of the measured $^{13}$N
to  $^{15}$O  neutrino  rates  in  the observation  window,  a  correction  we
introduce to  convert the total rate  to the rate for  $^{15}$O neutrinos.  In
principle this  is a measurable quantity:  because of the  lower endpoint, the
relative importance of $^{13}$N neutrinos  drops quickly with energy.  For the
$^{15}$O neutrinos,  60\% of the events  would reside in  bins between 0.8-1.0
MeV,  with the  remaining 40\%  between 1.0-1.3  MeV.  If  one looks  at total
events ($^{13}$N and $^{15}$O),  $^{13}$N neutrinos are responsible for $\sim$
19\%  of the  events between  0.8-1.0  MeV, but  only 1.0\%  of those  between
1.0-1.3 MeV, taking BPS08(AGS) SSM best-value fluxes.
Such  a bin analysis will have to  contend with the contribution
from  the line-source pep  neutrinos as  well, however,  so the  accuracy with
which $\alpha$ can be measured is certainly not clear.

However, an experimental determination is probably not required.  We can write
$\alpha$ as
\begin{equation}
\alpha  =   {\phi(^{13}\mathrm{N})  \over  \phi(^{15}\mathrm{O})}   {  \langle
\sigma^{B/S}(^{13}\mathrm{N},   \delta  m_{12}^2,\theta_{12})   \rangle  \over
\langle \sigma^{B/S}(^{15}\mathrm{O}, \delta m_{12}^2,\theta_{12}) \rangle}
\end{equation}
The cross section ratio can  be evaluated, yielding 0.086(1 $\pm$ 0.0036), when
the LMA oscillation  parameters are varied over the full  range allowed by the
KamLAND  combined analysis.   The  reason for  the  very small  error is  that
variations  in  these  parameters  tend  to  affect  the  two  cross  sections
identically: the 0.8-1.3  MeV event window is narrow,  and clearly differences
must  vanish in  the limit  of a  zero-width window.   The flux  ratio  at the
parameter point defined by the BPS08(AGS) SSM
best values is 1.40.  The changes in 
this ratio obtained  by varying SSM parameters can  be evaluated by procedures
similar to those leading to Eqs. (\ref{eq:prod4}) and (\ref{eq:prod4b}),
\begin{eqnarray}
\alpha  &=&  0.120   (1  \pm  .0036)  \left[{\phi^{SNO}(^{8}\mathrm{B})  \over
\phi^{SSM}(^{8}\mathrm{B})}\right]^{-0.229} \nonumber \\ &\times& \left[ 1 \pm
0.24\% (\mathrm{resid.~envir.}) \pm 2.0\% (\mathrm{nuclear}) \right] \nonumber
\\   &\times&  \left(   {X(^{12}\mathrm{C})   \over  X(^{12}\mathrm{C})_{SSM}}
\right)^{0.053}  \left(  {X(^{14}\mathrm{N})  \over  X(^{14}\mathrm{N})_{SSM}}
\right)^{-0.058} ~~
\label{eq:prod4c}
\end{eqnarray}
One  finds that  $\alpha$  is stable  under  reasonable parameter  variations:
changes  induced by the  nuclear physics  or core  temperature tend  to affect
these fluxes  in similar  ways.  This includes variations  in core
metals, the  quantity we hope  to constrain: adjustments in the C  or N
primordial abundance by 30\% produce changes at or below 1.5\%, that is, 0.120
(1 $\pm$ 0.015).   If the overall metalicity is changed,  keeping the C/N ratio
fixed, this becomes 0.1\%.  One  concludes from these exercises that the ratio
of  events in  the  0.8-1.3 MeV  window  from $^{13}$N  and $^{15}$O  neutrino
interactions to that from $^{15}$O neutrinos alone, is 1.120 $\pm$ 0.003, when
all sources of uncertainty are considered.

The final step is  to plug Eqs.~\ref{eq:fluxratio}~and~\ref{eq:fluxratio2}
into Eq.~\ref{eq:prod3a} to obtain
\begin{eqnarray}
 {R_{exp}^{B/S}(\mathrm{CN})   \over   R^{B/S}_{cal}(^{15}\mathrm{O},   \delta
m_{12}^2,\theta_{12}))  }  =  ~~~~~~~~~~~~~~~~~~~~~~~~~~~~~~~~~~~\nonumber  \\
(1.120     \pm     0.003)     \left[     {R^{SK}_{exp}(^8\mathrm{B})     \over
R^{SK}_{cal}(^8\mathrm{B},            \delta            m_{12}^2,\theta_{12})}
\right]^{0.828}~~~~~~~~~~   \nonumber   \\   \times   \left[   1   \pm   2.6\%
(\mathrm{resid.~envir.})  \pm 7.6\%  (\mathrm{nuclear})  \right] \nonumber  \\
\times    \left(     {X(^{12}\mathrm{C})    \over    X(^{12}\mathrm{C})_{SSM}}
\right)^{0.805}  \left(  {X(^{14}\mathrm{N})  \over  X(^{14}\mathrm{N})_{SSM}}
\right)^{0.199}.~~~~~~~~~
\label{eq:final}
\end{eqnarray}
The SK rate term is the  experimental ``thermometer'' we use to remove most of
the solar  model ``environmental'' uncertainty,  leaving in the next  term SSM
uncertainties  that   are  dominated  by  the  nuclear   physics.   But  these
uncertainties are,  in some sense, under  our control, and will  be reduced as
laboratory reaction measurements continue.   The last terms are the primordial
abundances we would  like to constrain.  The role of the  SSM in this equation
is to  define a set  of parameters  and thus a  set of reference  rates, about
which  we then  explore possible  variations.  Those  variations  generate the
environmental   and   nuclear   uncertainties   given  above,   according   to
Eq.~\ref{eq:prod3a}.

The   $R_{cal}$   factors   in   Eq.   (\ref{eq:final})   contain   additional
uncertainties, including one important one:
\begin{itemize} 
\item The shape of the normalized neutrino spectra: The $^{15}$O shape spectrum
is allowed,  and thus  accurately known.  The  $^8$B spectrum is  less certain
because the $\beta$ decay populates  a broad final-state resonance.  In the SK
analysis this spectrum  error is among those included  in the systematic error
budget,  so  it  should  not   be  counted  a  second  time.   It  contributes
\citep{hosaka}  at the  $\sim$ 1\%  level.  This  is a  laboratory astrophysics
uncertainty  that could  be lowered  by  improved measurements  of the  $^8$Be
resonance.
\item Uncertainties  in the  elastic scattering cross  section are  also small
($\sim$ 0.5\% \citealt{hosaka}), and furthermore tend to cancel between the two
normalizing cross sections in Eq.(\ref{eq:final}).
\item The principal uncertainty in  the cross section ratio is that associated
with neutrino  oscillations. Apart  from the dependence  on the  solar density
profile, one can  consider this to be another  type of laboratory uncertainty:
oscillation parameters  can and  will be further  constrained by a  variety of
accelerator and reactor experiments.   For example, KamLAND currently provides
our best constraint on $\delta m^2_{12}$.
\end{itemize}

The  LMA parameter  uncertainties  in Super-Kamiokande  and Borexino/SNO+  are
anti-correlated.  Most of the  low-energy $^{15}$N neutrinos do not experience
a level  crossing, residing instead  in a portion  of the MSW plane  where the
oscillations are close to the vacuum oscillation limit:
\begin{equation}
P_{\nu_e}(E_\nu) \rightarrow 1 - {1 \over 2} \sin{2 \theta_{12}}
\end{equation}
Thus  an increase  in  the  vacuum mixing  angle  $\theta_{12}$ decreases  the
$\nu_e$ survival  probability.  The higher energy $^8$B  neutrinos are largely
within  the MSW  triangle,  described  by an  adiabatic  level crossing.   The
limiting behavior for an adiabatic crossing is
\begin{equation}
P_{\nu_e}(E_\nu) \rightarrow {1 \over 2} (1 -\cos{2 \theta_{12}})
\end{equation}
so that an increase in $\theta_{12}$ increases the survival probability.  This
anti-correlation thus leads to larger effects in the ratio.

We have evaluated the impact of this uncertainty on Eq. (\ref{eq:final}), using
the  allowed  regions for  $\theta_{12}$  and  $\delta  m_{12}^2$ obtained  in
KamLAND's  combined analysis  \citep{kamland,kamland2},  $0.833 \lsim  \sin^2{2
\theta} \lsim  0.906$ and $ 7.38  \times 10^{-5}$ eV$^2  \lsim \delta m_{12}^2
\lsim 7.80 \times 10^{-5}$ eV$^2$.  This yields
\begin{eqnarray}
{R^{B/S}_{cal}(^{15}\mathrm{O},     \delta     m_{12}^2,\theta_{12}))    \over
 R^{SK}_{cal}(^8\mathrm{B},     \delta     m_{12}^2,\theta_{12})^{0.825}}    =
 ~~~~~~~~~~~~ \nonumber \\ (1 \pm 0.049)\left[ {R^{B/S}_{cal}(^{15}\mathrm{O},
 \delta   m_{12}^2,\theta_{12}))   \over   R^{SK}_{cal}(^8\mathrm{B},   \delta
 m_{12}^2,\theta_{12})^{0.825}} \right]^{BV}
\end{eqnarray}
where $BV$ denotes the best-value ratio.

Thus  the overall  uncertainty budget  in Eq.  {\ref{eq:final}}  appears quite
favorable,  with the SK  ``thermometer'' contributing  at 3\%,  residual solar
environmental uncertainties at 1\%,  and LMA parameter uncertainties at 4.9\%.
The  largest  of the  errors  is  that  from nuclear  S-factor  uncertainties,
currently 7.6\%.  The overall  uncertainty in the ``theoretical'' relationship
between a future SNO+ or Borexino CN-neutrino flux and core C/N metals is thus
about 9.6\%.  As  the nuclear physics uncertainty dominates  the analysis, one
would expect this relationship to become more precise when ongoing analyses of
the full data set for $^{14}$N(p,$\gamma$) are completed.  An appropriate goal
would  be 3.5\%  in this  S-factor, a  30\% improvement.   The  uncertainty in
$^{14}$N(p,$\gamma$)  would  no  longer  dominate the  nuclear  physics  error
budget, but instead would be comparable to the contributions from S$_{33}$ and
S$_{34}$.  However, the current 9.6\% uncertainty is not a bad starting point,
as first-generation CN-cycle neutrino experiments are expected to measure this
flux to an accuracy of about  10\%.  That is, the theoretical uncertainty will
not dominate the experimental uncertainty.

\section{Future Experiments and Summary}

One of the main motivations for  this paper is the development of new detector
ideas  that  might  allow   a  high-statistics  measurement  of  the  CN-cycle
neutrinos.    In   particular,   detectors   based  on   ultra-clean   organic
scintillation liquids  have, at  least in principle,  the potential to  make a
high-sensitivity real-time measurement of the CN-cycle neutrinos.

The  Borexino collaboration has  investigated this  possibility \citep{franco}.
Borexino, currently operating in Gran Sasso, has a 300-ton liquid scintillator
target housed in a 8.5m spherical  nylon membrane and shielded by a kiloton of
buffer fluid.  Events  producing light within the detector  are detected by an
array  of 2200  photomultiplier tubes.   The inner  100 tons  of  the detector
comprise the  fiducial volume.   The events come  from elastic  scattering off
electrons, a  reaction that  is sensitive to  both electron and  (with reduced
sensitivity) heavy-flavor neutrinos.

Borexino is primarily focused on detecting neutrinos from  the pp chain,
specifically the 862 keV line neutrinos from electron capture on
$^7$Be (the 90\% branch).  However, as mentioned previously, the
collaboration has proposed using the
detection window of (0.8-1.3) MeV to pick up
contributions from the pep line source (1.442 MeV) and the $^{13}$N and
$^{15}$O $\beta$ decay sources. 

The primary  obstacle to such a measurement  by Borexino is the  {\it in situ}
cosmogenic  production of  $^{11}$C,  a $\beta^+$  source.  High-energy  muons
produced by primary cosmic ray interactions in the atmosphere can penetrate to
great  depths, producing  $^{11}$C  by  knocking a  neutron  out of  $^{12}$C.
Borexino is located  in Gran Sasso, which has a depth  of about 3.1 kilometers
of water equivalent (kmwe), when converted to the equivalent depth below a flat
surface \citep{meihime}.  The $^{11}$C production is still significant at this
depth: an initial estimate of the associated background of 7.5 c/d/100 tons in
Borexino was recently confirmed by a direct measurement \citep{back}, yielding 
13.0 $\pm$  2.6 (stat) $\pm$ 1.4  (sys) c/d/100 tons.  This  exceeds the solar
neutrino signal  in the window of  interest.  Thus some means  of vetoing this
background must be introduced.  This vetoing is nontrivial because of the long
mean lifetime of $^{11}$C, 29.4 minutes:  a simple cut based on the muon would
thus not be feasible.

The collaboration  has proposed  that a successful  veto might be  possible by
exploiting a triple coincidence, the  initiating muon, the prompt capture of a
neutron on protons in the  scintillator fluid, and the delayed $\beta^+$ event
(\citealt{franco};  Deutsch,  private communication).   This  would allow  the
  experimenters to cut out a 
spherical  volume defined  by  the neutron  capture  vertex, rejecting  events
within  that  volume for  a  time $\Delta$t  large  compared  to the  $^{11}$C
lifetime.    The   simulations   performed   by  Borexino   suggest   that   a
signal/background  ratio of  1.2  could  be achieved  with  20\% deadtime.   A
CN-cycle neutrino  (and pep neutrino) flux  measurement could then  be made by
subtraction of this background.

An alternative  to this  approach would be  to place  such a detector  at very
great depth.  This  possibility has been discussed by  the SNO+ collaboration,
which has  proposed placing a  one-kiloton scintillator experiment  in SNOLab,
using the  cavity that was originally  excavated for SNO  \citep{chen}.  Such a
detector  could be  used for  detecting $^7$Be,  pep, and  CN-cycle neutrinos,
geoneutrinos, and double beta decay.  The proposed volume is about a factor of
three greater than that of Borexino.  The great advantage of SNO+ would be its
depth, 6.0 kmwe,  and consequently its much lower  cosmic ray muon background.
The additional  3.0 kmwe, relative to  Gran Sasso, provides about  a factor of
$\sim$  70 additional  attenuation  in the  muon  flux, so  that the  expected
$^{11}$C production would be reduced to $\sim$ 0.1 c/d/100 tons, a few percent
of the expect CN neutrino signal.

\begin{figure*}
\includegraphics[]{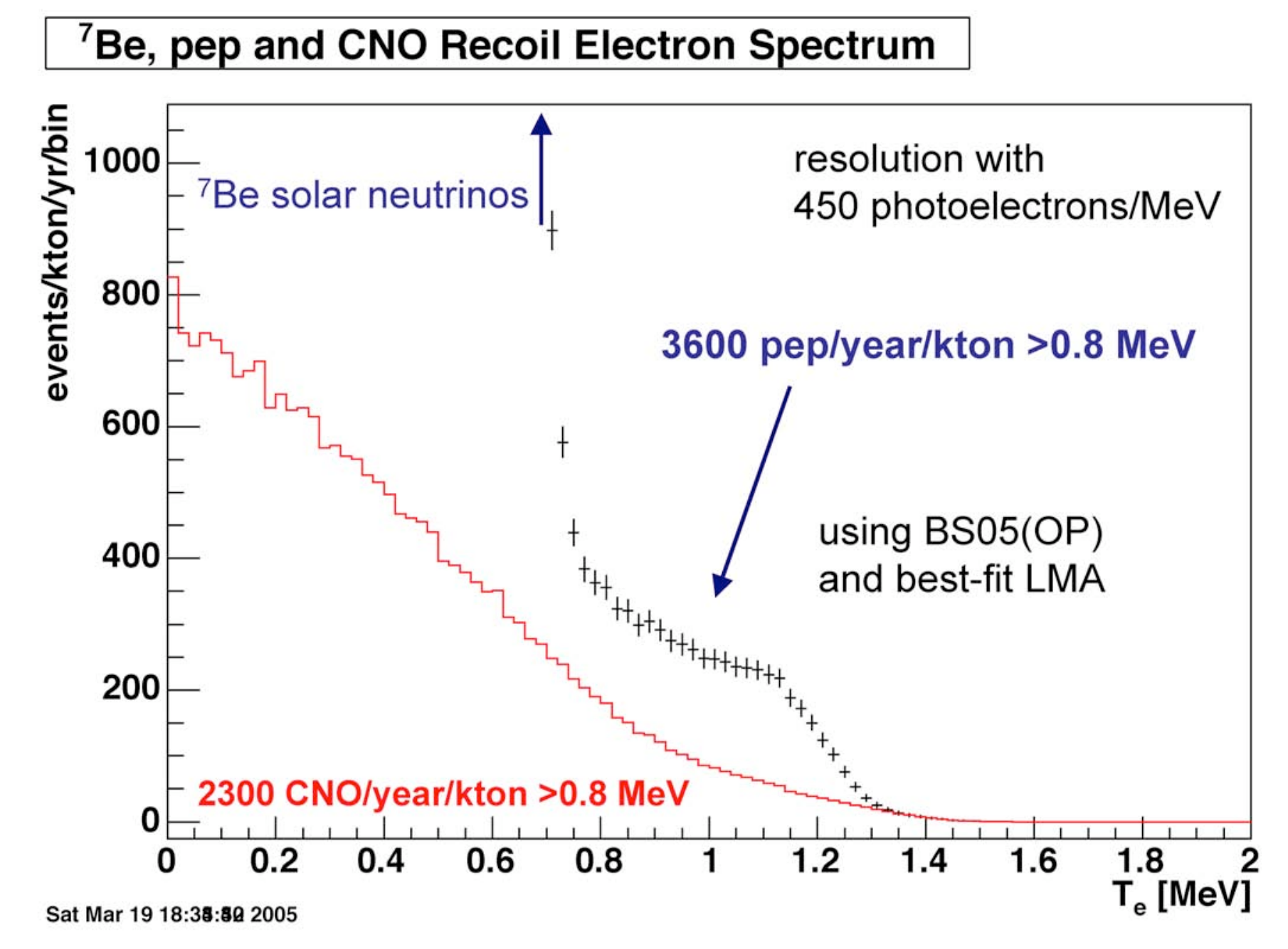}
\caption{A simulation of events expected in SNO+, the proposed SNOLab
experiment to measure low-energy solar and other neutrino sources.
This figure is due to M. C. Chen \citep{chen}. \label{fig:CNO_SNOPlus}}
\end{figure*}

Figure \ref{fig:CNO_SNOPlus} shows a simulation of the expected SNO+ response,
performed by the experimenters (Chen, private communication).  Note that
the simulation is based on  the  BS05(OP) SSM and the best-fit LMA solution 
to  the solar  neutrino problem, rather than the updated BPS08(AGS) used in this
paper.  The CN-neutrino  event rate  for  an energy
window  above 0.8 MeV  was found  to be  2300 counts/year.   The experimenters
concluded that  SNO+ could  determine the CN-neutrino  rate to an  accuracy of
approximately 10\%, after three years of running \citep{chen}.  This accuracy
is  the  appropriate  goal  for  such  a  first-generation  CN-cycle  neutrino
measurement, as it  would approach the accuracy with which  that flux could be
related theoretically to  the Sun's primordial core C and  N abundances, as we
have argued in this paper.

In  this  paper  we  have  suggested  a possible  strategy  for  using  future
Borexino/SNO+ CN-neutrino  measurements as  a test of  the primordial C  and N
abundances in the solar core.  The approach is based on using Super-Kamiokande
as a solar thermometer, to  largely eliminate other SSM uncertainties, so that
a clean  relationship between  these abundances and  CN neutrino rates  can be
made.  SNO,  KamLAND, and other  neutrino oscillation experiments are  used in
the  analysis   to  constrain  LMA  oscillation  parameters.    We  derived  a
relationship  where the dominant  linear dependence  on C  and N  remains, but
other solar-model dependences are  largely eliminated.  This approach exploits
the logarithmic derivatives  that have been previously calculated  for the SSM
(especially those for the separate metal abundances) that define the impact of
SSM  parameter  variations, supplemented by Monte Carlo calculations (which
treat explicitly any correlations that may exist among the parameter variations).
Although  the  relationship  is  derived in  the
context of  the SSM, we suspect  that it remains  valid for a larger  group of
models, e.g., those where SSM parameters are varied well beyond their accepted
SSM uncertainties: many investigators have  shown that the $^8$B neutrino flux
remains a reliable thermometer, even  when such large SSM parameter variations
are made.

We have found that the factors that limit the accuracy of Eq. (\ref{eq:final})
are first, uncertainties in nuclear  cross sections ($\sim$ 7.6\%) and second,
uncertainties  in LMA oscillation  parameters ($\sim$  4.9\%).  Both  of these
uncertainties can and will be  reduced in future laboratory measurements.  One
goal should be  the reduction, eventually, to the level  of uncertainty of the
SK thermometer, $\sim$ 3\%.

In  summary,  it appears  possible  to  use  future experiments  sensitive  to
CN-cycle neutrinos to constrain C and  N content of the Sun's primordial core.
This would test an important assumption made in the SSM, that the zero-age Sun
was  homogeneous,  with  a  core  metalicity  identical  to  that  of  today's
photosphere.

Such a measurement  would also address a significant  controversy, that recent
3D models of photospheric absorption lines  have led to lower estimates of the
abundances of the volatile metals.  These new analyses appear to be on a solid
foundation,  substantially  improving  absorption  line  systematics  and  the
consistency between the  Sun and other similar stars in  the local group.  Yet
they also significantly alter SSM predictions  of the sound speed in the upper
part  of the  Sun's radiative  zone, so  that  the SSM  is no  longer in  good
agreement with  constraints imposed by  helioseismology.  For this  reason, an
independent measurement of the C and  N abundances in the Sun's radiative core
would be of great interest.

This measurement would also place  an important experimental constraint on the
evolutionary history of the Sun.  While  the argument for a homogeneous Sun at
the end of the pre-main-sequence  Hayashi phase appears credible, once the Sun
begins to form a radiative core, there are no subsequent SSM epochs that would
allow mixing of the full Sun.   Thus is principle any anomaly in the accretion
of  metals onto the  Sun, either  during the  main sequence  or in  the Henyey
phase, could produce chemical inhomogeneities.   Such a scenario was the basis
of the low-Z model, one of the proposed solar solutions to the puzzle posed by
the chlorine experiment.

Unlike the low-Z  model, a naive attempt to accommodate  both the sound speeds
required  by helioseismology and  the new  photospheric abundances  requires a
convective zone  depleted in  metals.  We have  noted that the  solar system's
primary reservoir for  metals, the gaseous giant planets,  are thought to have
formed  late relative to the evolution of the proto-Sun,
incorporating an excess of metal  estimated at 40-90 M$_\oplus$.  This mass is
similar to the deficit of metals in the convective zone, were one to interpret
the helioseismology/photospheric abundance discrepancy  in the most naive way.
The  raises an  provocative question:  is it  possible that  the  process that
concentrated  metals in the  gaseous giants  also produced  a large  volume of
metal-depleted  gas that  subsequently was  accreted onto  the  Sun's surface?
While the  suggestion of  a common chemical  mechanism linking  the convection
zone  and  the gaseous  giants  is  speculative, we  think  this  is one  more
motivation for  exploiting the CN neutrinos  as a quantitative  probe of solar
core metalicity.

\acknowledgments

We thank  M. Chen,  P. Goldreich, K.  Snover, Y.  Suzuki, and M.  Wiescher for
discussions.   This  work was  supported  in part  by  the  Office of  Nuclear
Physics,  US Department  of  Energy, under  grant  DE-FG02-00ER-41132. AMS  is
supported  by the  IAS through  a John  Bahcall Fellowship  and NSF grant
PHY-0503584. 

\bibliographystyle{apj}

\clearpage



\clearpage


\clearpage

\clearpage

\clearpage

\begin{deluxetable}{lcccccccccc}
\tabletypesize{\small}
\tablewidth{0pt}
\tablecaption{Partial  derivatives  $\alpha(i,j)$  of  neutrino  fluxes  with
    respect  to  solar environmental  and  nuclear  cross section  parameters.
    \label{table:one}} 
\tablehead{ & \multicolumn{4}{c}{Environmental $\beta_j$} &
\multicolumn{6}{c}{Nuclear $\beta_j$}} 
\startdata
Source  & $L_\odot$  &  Opacity &  Age  & Diffusion~&  $S_{11}$  & $S_{33}$  &
$S_{34}$ & $S_{17}$ & $S_{e7}$ & $S_{114}$ \\ 
\hline
$\phi$($^8$B) & 7.16 & 2.70 & 1.38 & 0.28 & -2.73 & -0.43 & 0.85 & 1.0 & -1.0 &
-0.020 \\ 
$\phi$($^{13}$N) & 4.40 & 1.43 & 0.86 & 0.34 & -2.09 & 0.025  & -0.053 & 0.0 &
0.0 & 0.71 \\ 
$\phi$($^{13}$N)/$\phi$($^8$B)$^{0.599}$ & 0.11 & -0.19  & 0.03 & 0.17 & -0.45
& 0.28 & -0.56 & -0.60 & 0.60 & 0.72 \\ 
$\phi$($^{15}$O) & 6.00 & 2.06 & 1.34 & 0.39 & -2.95 & 0.018 & -0.041 & 0.0 &
0.0 & 1.00 \\ 
$\phi$($^{15}$O)/$\phi$($^8$B)$^{0.828}$ & 0.07 & -0.18  & 0.20 & 0.16 & -0.69
& 0.37 & -0.74 & -0.83 & 0.83 & 1.02 \\ 
\enddata
\tablecomments{Table   entries  are   the   logarithmic  partial   derivatives
$\alpha(i,j)$  of the  solar  neutrino  fluxes $\phi_i$  with  respect to  the
indicated solar  model parameter $\beta_j$,  taken about the SSM  best values.
All fluxes are in units of  their SSM best values, and thus are dimensionless.
The derivatives, taken from \citet{bps08}, are for the SSM 
BPS08(AGS), which  employs the AGS abundances.  The  two flux ratios were
determined   for  a   $\phi(^8$B)   exponent  that   minimizes  the   residual
environmental error  in the prediction, including  the environmental variables
here and  in Table \ref{table:two}.  As  explained in the text,  that error is
weighted  according  to  the  uncertainties in  the  environmental  parameters
$\beta_j$, given in Table~\ref{table:three}. 
} 
\end{deluxetable}

\clearpage

\begin{deluxetable}{lccccccccc}
\tabletypesize{\small}
\tablewidth{0pt}
\tablecaption{Partial  derivatives  $\alpha(i,j)$  of  neutrino  fluxes  with
    respect    to   fractional    abundances   of    the    primordial   heavy
    elements. \label{table:two}} 
\tablehead{       &      \multicolumn{2}{c}{C,       N       $\beta_j$}      &
  \multicolumn{6}{c}{Environment Abundance $\beta_j$}} 
\startdata
Source & C& N & O & Ne & Mg & Si & S & Ar & Fe \\
\hline
$\phi$($^8$B) & 0.027 & 0.001 & 0.107 & 0.071 & 0.112 & 0.210 & 0.145 & 0.017
& 0.520 \\ 
$\phi$($^{13}$N) &  0.874 & 0.142 &  0.044 & 0.030 &  0.054 & 0.110  & 0.080 &
0.010 & 0.268 \\ 
$\phi$($^{13}$N)/$\phi$($^8$B)$^{0.599}$ &  0.858 & 0.141 & -0.020  & -0.013 &
-0.013 & -0.016 & -0.007 & 0.000 & -0.043 \\ 
$\phi$($^{15}$O) &  0.827 & 0.200 &  0.071 & 0.047 &  0.080 & 0.158  & 0.113 &
0.013 & 0.393 \\ 
$\phi$($^{15}$O)/$\phi$($^8$B)$^{0.828}$ &  0.805 & 0.199 & -0.018  & -0.012 &
-0.013 & -0.016 & -0.007 & -0.001 & -0.038 \\ 
\enddata
\tablecomments{Heavy  elements are  divided into  ``environmental''  metals --
those which  primarily influence the solar  core through their  effects on the
opacity  and thus  the  core temperature  -- and  C  and N,  which govern  the
production  of  $^{13}$N  and  $^{15}$O  solar  neutrinos  and  which  can  be
determined,  in   principle,  from  measurements  of   these  fluxes.  Results
correspond to the BPS08(AGS) model \citep{bps08}.} 
\end{deluxetable}

\clearpage

\begin{deluxetable}{lccccc}
\tabletypesize{\small} 
\tablewidth{0pt} 
\tablecaption{Estimated  1$\sigma$  uncertainties  in  solar and  nuclear  SSM
parameters,  taken  from  Bahcall,  Serenelli,  and  Basu  \cite{bsb_mc}  and
Fiorentini  and Ricci  \cite{fr},  and their  influence  on flux  predictions,
computed  from   the  partial  derivatives  of   Table  \ref{table:one}.   The
experimental value for $S_{17}$ is taken from a series of recent measurements:
this $S$-factor and $S_{114}$ are discussed in the text. \label{table:three}}
\tablehead{$\beta_j$  &   Value  &  $\Delta   \beta_j/\beta_j$(\%)  &  $\Delta
  \phi(^8$B)/$\phi(^8$B)(\%)  &   $\Delta  \phi(^{13}$N)/$\phi(^{13}$N)(\%)  &
  $\Delta \phi(^{15}$O)/$\phi(^{15}$O) (\%)} 
\startdata 
L$_\odot$ & 3.842 $\times$ 10$^{33}$ ergs/s & 0.4 & 2.9 & 1.8 & 2.4 \\ 
Opacity & 1.0 & 2.5 & 6.9 & 3.6 & 5.2 \\ 
Age & 4.57 b.y. & 0.44 & 0.61 & 0.38 & 0.59 \\ 
Diffusion & 1.0 & 15.0 & 4.0 & 4.9 & 5.7 \\ 
p+p & 3.94 $\times$ 10$^{-25}$ MeV b & 0.4 & 1.1 & 0.84 & 1.2 \\ 
$^3$He+$^3$He & 5.4 MeV b & 6.0 & 2.5 & 0.15 & 0.10 \\ 
$^3$He+$^4$He & 0.564 MeV b & 4.9 & 4.1 & 0.25 & 0.20 \\ 
p+$^7$Be & 20.6 eV b & 3.8 & 3.8 & 0.0 & 0.0 \\ 
e+$^7$Be & & 2.0 & 2.0 & 0.0 & 0.0 \\ 
p+$^{14}$N & 1.61 keV b & 5.0 & 0.1 & 3.5 & 5.0 \\ 
\enddata
\end{deluxetable}

\clearpage

\begin{deluxetable}{lcccc}
\tabletypesize{\small} 
\tablewidth{0pt} 
\tablecaption{Estimated 1$\sigma$  historical (``conservative'') uncertainties
in  AGS abundances,  as defined  in  \cite{bs05}.   The
corresponding  uncertainties in  the  neutrino fluxes  are  computed from  the
partial derivatives of Table~\ref{table:two}.\label{table:four}
}
\tablehead{$\beta_j$     &    $\Delta    \beta_j/\beta_j$(\%)     &    $\Delta
  \phi(^8$B)/$\phi(^8$B)(\%)  &   $\Delta  \phi(^{13}$N)/$\phi(^{13}$N)(\%)  &
  $\Delta \phi(^{15}$O)/$\phi(^{15}$O)(\%)
}
\startdata
C & 29.7 & 0.70 & 25.5 & 24.0 \\ 
N & 32.0 & 0.03 & 4.0 & 5.7 \\
O & 38.7 & 3.6 & 1.4 & 2.3 \\
Ne & 53.9 & 3.1 & 1.3 & 2.0 \\
Mg & 11.5 & 1.2 & 0.59 & 0.87 \\
Si & 11.5 & 2.3 & 1.2 & 1.7 \\
S & 9.2 & 1.3 & 0.71 & 1.0 \\
Ar & 49.6 & 0.69 & 0.40 & 0.53 \\
Fe & 11.5 & 5.8 & 3.0 & 4.4 \\ 
\enddata
\end{deluxetable}

\end{document}